\documentclass[twocolumn,superscriptaddress,floatfix,
longbibliography,aps,prb,preprintnumbers,10pt,groupedaddress]{revtex4-2}
\usepackage{titlesec}

\usepackage{graphicx}
\usepackage{amssymb}
\usepackage{bm}
\usepackage{graphicx,color}

\usepackage[urlcolor=blue,colorlinks=true,citecolor=blue,linkcolor=blue,pdfstartview={FitH},bookmarks=false]{hyperref}
\newcommand{\sgn}{{\rm sgn}}
\newcommand{\bet}{\mbox{\boldmath{$\beta$}}}
\newcommand{\sig}{\mbox{\boldmath{$\sigma$}}}
\newcommand{\kap}{\mbox{\boldmath{$\kappa$}}}
\newcommand{\nabb}{\mbox{\boldmath{$\nabla$}}}

\begin{document}

\title{Instability of the ferromagnetic quantum critical point in strongly interacting 2D and 3D electron gases
with {arbitrary} spin-orbit splitting} 

\date{\today}
\author{Dmitry Miserev,$^{1\ast}$ Daniel Loss,$^{1}$ and Jelena Klinovaja$^{1}$}
\affiliation{$^{1}$Department of Physics, University of Basel, \\
	Klingelbergstrasse 82, CH-4056 Basel, Switzerland\\
}

\begin{abstract}
In this work we revisit itinerant ferromagnetism 
in 2D and 3D electron gases with arbitrary spin-orbit splitting 
and strong electron-electron interaction.
We identify the resonant scattering processes
close to the Fermi surface
that are responsible for the instability of 
the ferromagnetic quantum critical point at low temperatures.
In contrast to previous theoretical studies,
we show that such processes cannot be
fully suppressed even in presence of arbitrary 
spin-orbit splitting. 
A fully self-consistent non-perturbative treatment 
of the electron-electron interaction close to the phase transition
shows that these resonant processes
always destabilize the ferromagnetic quantum critical point
and lead to a first-order phase transition.
Characteristic signatures of these processes 
can be measured via the non-analytic dependence of the spin susceptibility on
magnetic field both far away 
or close to the phase transition.
\end{abstract}

\maketitle

\section{Introduction}

Itinerant ferromagnetism in two-dimensional (2D)
and three-dimensional (3D) electron gas
has been observed in various materials,
such as  
manganese perovskites \cite{ramirez,coey,ghosh,kim},
transition-metal-doped semiconductors \cite{dietl,ohno,ohno2},
monolayers of transition metal dichalcogenides
\cite{tongay,bonilla,roch1,roch2,huang},
and many others \cite{bozorth,boer,takagi,nakabayashi,deiseroth,almamari,xu,walter}.
The physical mechanisms leading to the ferromagnetic ground state
depend strongly on the materials.
Doping by  transition metals
results in strong interaction 
between the itinerant spins and 
the magnetic moments of the transition metal ions,
the mechanism known as  double exchange or Zener mechanism \cite{zener,alexandrov,verges}.
In this case the ferromagnetism is not intrinsic 
but rather induced by the magnetic moments of the dopants.
In contrast, in this work, we are interested in the itinerant ferromagnetism
emerging from strong electron-electron interactions
between the delocalized charge carriers.
This mechanism is often referred to as  Stoner mechanism \cite{stoner,stoner2}.

In the original work of Stoner \cite{stoner}
the phase transition to the ferromagnetic phase is of second order, i.e. continuous.
The ferromagnetic quantum critical point 
(FQCP) 
of the spin-degenerate electron gas is analyzed in the literature via  
the effective Ginzburg-Landau-Wilson theory \cite{ginzburg,wilson,hertz,millis,sachdevcrit}
describing 
the fluctuating magnetic order parameter.
However, this theory relies on 
the analyticity of the effective Lagrangian \cite{millis,sachdevcrit}
which does not hold for 
interacting 2D and 3D electron gases \cite{vojta}.
The negative non-analytic corrections,
originating from
the resonant backscattering 
of itinerant electrons 
close to the spin-degenerate Fermi surface,
emerge already in second order perturbation in the electron-electron interaction
\cite{vojta,belitz1,belitz2,chubukov1,chubukov2,belitz3,belitz4,lohneysen,belitzrev}.
If these non-analyticities survive in a small vicinity of the ferromagnetic quantum phase transition (FQPT), they
destabilize the FQCP at zero temperature 
and lead to a first-order FQPT
in 2D and 3D electron gases \cite{belitz1,belitz2,chubukov1,chubukov2,belitz3,belitz4,lohneysen,belitzrev}.
This phenomenon is an example of 
the fluctuation-induced first-order transition
first predicted in high energy physics \cite{coleman}.
In condensed matter, this effect is responsible for weak first-order metal-superconductor and smectic-nematic phase transitions \cite{halperin}.

The problem with FQPTs in clean metals is that it happens at very strong electron-electron interaction 
where the perturbative approach \cite{vojta,belitz1,belitz2,chubukov1,chubukov2,belitz3,belitz4,lohneysen,belitzrev} 
is no longer valid.
It has been pointed out in the literature that higher order scattering processes  
in the spin-degenerate electron gas
\cite{saha,maslov}
may change the sign of the non-analytic terms making them irrelevant in the infrared limit
and thus, stabilizing the FQCP.
So far, the greatest advances in understanding the strongly interacting regime are attributed to the effective spin-fermion model \cite{abanov}
where the collective spin excitations in strongly interacting electron gases 
are coupled to the electron spin.
The negative non-analyticities calculated within this model remain relevant,
although strongly suppressed near the FQPT \cite{saha,maslov}.

Numerical simulations of the low-density spin-degenerate 2D electron gas (2DEG) confirm a first-order FQPT
in the liquid phase
with further transition to the Wigner solid
at even lower densities
\cite{tanatar,rapisarda,varsano,attaccalite}.
The situation is 
less definite
in a 3D electron gas (3DEG)
where various advanced numerical techniques 
predict either a first- or second-order FQPT
depending on the numerical scheme
\cite{adler,ortiz,zong,loos,gross,holzmann}.
This disagreement between different numerical results for 3DEGs
might follow from the much weaker character of the non-analytic terms destabilizing the FQCP 
compared to the 2D case \cite{vojta}.

The problem becomes even more complicated if a spin-orbit (SO) splitting 
of the Fermi surface is present.
The main effect of the SO splitting is the spin symmetry breaking 
restricting possible directions of the net magnetization in the magnetically ordered phase \cite{zak1,zak2}.
In particular, the Rashba SO splitting in 2DEGs restricts possible net magnetization directions in the 2DEG plane \cite{zak1,zak2}.
So far, SO splitting is considered in the literature as a promising intrinsic mechanism cutting the non-analyticity and stabilizing the FQCP in the interacting electron gas
\cite{kirk}.

In this work, we consider the general case of a $D$-dimensional electron gas, $D > 1$, with arbitrary SO splitting and identify the resonant scattering processes close to the Fermi surface
which result in the non-analytic corrections with respect to the magnetization.
Our results are in perfect agreement with the previously considered case of the Rashba 2DEG
\cite{zak1,zak2}.
However, we show that even arbitrary SO splitting is not able to cut negative non-analytic corrections in 2DEGs and 3DEGs.
Thus, in contrast to Ref.~\cite{kirk}, we find that SO splitting cannot be considered as a possible intrinsic mechanism stabilizing FQCP in a uniform electron gas.

In this work we apply the dimensional reduction of the electron Green function which we developed earlier in Ref.~\cite{miserev}.
This procedure allows us to reduce $D$ spatial dimensions to a single effective spatial dimension
and significantly simplifies the derivation of the non-analytic corrections in the perturbative regime for arbitrary SO splitting.
We confirm the validity of our approach by comparison with known results \cite{maslov,zak1,zak2}.
In order to access the strongly interacting regime, we treat the resonant scattering processes near the Fermi surface
within the self-consistent Born approximation
and solve it in the limit of strong interaction.
Within this approach, we find the non-Fermi liquid electron Green function which differs significantly from the Green function calculated within the effective spin-fermion model \cite{saha,maslov}.
Within our model, the non-analyticities are strongly enhanced close to the FQPT and remain negative at arbitrary SO splitting.
Thus, we conclude that the FQCP in strongly interacting 2DEGs and 3DEGs is intrinsically unstable.

In order to test our theoretical model experimentally, we suggest to measure the spin susceptibility in the paramagnetic phase close to the FQPT.
According to our predictions,
the spin susceptibility $\chi_{ij} (\bm B)$ close to the FQPT takes the form 
$\chi_{ij} (\bm B) - \chi_{ij} (0) \propto |B|^{\frac{D - 1}{2}}$ modulo powers of $\ln(E_F/|B|)$,
while the spin-fermion model predicts 
a much weaker scaling:
$\chi_{ij} (\bm B) - \chi_{ij} (0) \propto |B|^{\frac{3}{2}}$ for 2DEGs
and $\chi_{ij} (\bm B) - \chi_{ij} (0) \propto |B|^2 \ln \ln (E_F/|B|)$ for 3DEGs \cite{maslov},
where $E_F$ is the Fermi energy, $|B|$ is measured in  units of energy.
In the presence of SO splitting we also predict a
non-trivial tensor structure 
of $\chi_{ij} (\bm B) - \chi_{ij} (0)$,
which can also be used to identify the structure of the SO coupling.
The candidate materials for experiments
are the pressure-tuned 3D metals ZrZn$_2$ \cite{uhlarz}, UGe$_2$ \cite{taufour}, 
2D AlAs quantum wells \cite{hossain}
and many more \cite{brando}.

The paper is organized as follows.
In Sec. II we introduce 
the non-interacting electron gas in $D > 1$ spatial dimensions
with arbitrary spin splitting.
In Sec. III
we derive the asymptotics of the free electron Green function at large 
imaginary time $\tau \gg 1/E_F$ and large distance $r \gg \lambda_F$,
where $E_F$ is the Fermi energy, 
$\lambda_F$ is the Fermi wavelength.
In Sec. IV we use  second order perturbation theory
to derive the non-analytic correction
to the thermodynamic potential $\Omega$
with respect to the arbitrary spin splitting.
In Sec. V 
we calculate the electron Green function
in the limit of strong electron-electron interaction
and find a sector in the phase space 
where the Green function is non-Fermi-liquid-like.
In Sec. VI 
we calculate the thermodynamic potential 
$\Omega$
in the regime of strong interaction
and find that 
the non-analytic corrections are negative
and parametrically larger than the 
ones predicted in the weakly interacting limit.
In Sec. VII we derive the 
non-analytic corrections to the spin susceptibility
both far away and close to the FQPT.
Conclusions are given in Sec. VIII.
Some technical details are deferred to the Appendix.
\\

\section{Non-interacting electron gas with arbitrary spin splitting}

In this section we consider a
non-interacting
single-valley
electron gas in $D > 1$
spatial dimensions with arbitrary spin splitting.
The case of $D = 1$ is not included in this paper 
due to the Luttinger liquid instability of one-dimensional Fermi liquids with respect to arbitrarily small interactions \cite{luttinger}.
The electron gas is described by the following single-particle Hamiltonian:
\begin{eqnarray}
&& H_0 = \frac{p^2}{2 m} - E_F - \sig \cdot \bet(\bm p) , \label{H0}
\end{eqnarray}
where $\bm p$ is a $D$-dimensional momentum,
$m$ the effective mass,
$E_F$ the Fermi energy,
$\bet(\bm p)$ the spin splitting,
$\sig = (\sigma_x, \sigma_y, \sigma_z)$ the Pauli matrices.
The spin splitting is considered small compared to the Fermi energy:
\begin{eqnarray}
&& \beta(\bm p) \equiv |\bet(\bm p)| \ll E_F , \label{small}
\end{eqnarray}
but otherwise arbitrary.
Therefore,
the spin splitting 
close to the Fermi surface
can be parametrized by the
unit vector $\bm n_{\bm p} = \bm p / p$
along the momentum $\bm p$:
\begin{eqnarray}
&& \bet (\bm p) \approx \bet(\bm n_{\bm p}), \,\,\,\, \bm n_{\bm p} = \frac{\bm p}{p}, \,\,\,\, p = k_F = \sqrt{2 m E_F} .
\end{eqnarray}
Here we introduced $k_F$ as the Fermi momentum at zero spin splitting
$\bet (\bm p) = 0$.

The eigenvectors $|\sigma,\bm n_{\bm p}\rangle$ of the Hamiltonian $H_0$
correspond to the eigenvectors of the operator 
$\sig \cdot \bet(\bm n_{\bm p})$:
\begin{eqnarray}
&& \sig \cdot \bet (\bm n_{\bm p})  |\sigma, \bm n_{\bm p}\rangle = \sigma \beta(\bm n_{\bm p}) |\sigma, \bm n_{\bm p}\rangle ,  \label{eigenvectors}
\end{eqnarray}
where $\sigma = \pm 1$ and $\beta(\bm n_{\bm p}) = |\bet(\bm n_{\bm p})|$. 
The explicit form of the spinors is given by
\begin{equation}
|\sigma, \bm n_{\bm p}\rangle = \frac{\left(\beta_-(\bm n_{\bm p}) , \,
	\sigma \beta(\bm n_{\bm p}) - \beta_z(\bm n_{\bm p})\right)^T}{\sqrt{2 \beta(\bm n_{\bm p}) \left[\beta(\bm n_{\bm p}) - \sigma \beta_z (\bm n_{\bm p})\right]}} , \label{psip}
\end{equation}
where the superscript $^T$ means transposition,
$\beta_\pm (\bm n_{\bm p}) = \beta_x (\bm n_{\bm p}) \pm i \beta_y (\bm n_{\bm p})$. 
Two spinors with the same $\bm n_{\bm p}$ and opposite $\sigma$ are orthogonal:
\begin{eqnarray}
&& \langle +, \bm n_{\bm p}|-, \bm n_{\bm p}\rangle = 0 . \label{orthogonality}
\end{eqnarray}
This forbids the forward scattering 
between the bands 
with opposite band index $\sigma$.

In this paper we need the backscattering matrix elements:
\begin{eqnarray}
&& M_{\sigma\sigma'}(\bm n_{\bm p}) = \langle\sigma, \bm n_{\bm p} |\sigma', -\bm n_{\bm p}\rangle . \label{Msig}
\end{eqnarray}
Using Eq.~(\ref{psip}), we find the matrix elements explicitly:
\begin{widetext}
\begin{eqnarray}
&& M_{\sigma \sigma'}(\bm n_{\bm p}) = \frac{\beta_+(\bm n_{\bm p})\beta_-(-\bm n_{\bm p}) + \sigma \sigma' \left[\beta(\bm n_{\bm p})-\sigma \beta_z(\bm n_{\bm p})\right]
	\left[\beta(-\bm n_{\bm p})-\sigma' \beta_z(-\bm n_{\bm p})\right]}{\sqrt{4 \beta(\bm n_{\bm p}) \beta(-\bm n_{\bm p})\left[\beta(\bm n_{\bm p})-\sigma \beta_z(\bm n_{\bm p})\right]
\left[\beta(-\bm n_{\bm p})-\sigma' \beta_z(-\bm n_{\bm p})\right]}} . \label{mss}
\end{eqnarray}
\end{widetext}
The spin splitting $\bet (\bm n_{\bm p})$ results in two 
Fermi surfaces 
labeled by $\sigma = \pm 1$
with the Fermi momenta being
dependent on $\bm n_{\bm p}$:
\begin{eqnarray}
&& k_\sigma (\bm n_{\bm p}) = \sqrt{2 m \left(E_F + \sigma \beta (\bm n_{\bm p})\right)} \approx k_F + \sigma \frac{\beta(\bm n_{\bm p})}{v_F} , \label{ksig}
\end{eqnarray}
where $v_F = k_F/m$ is the Fermi velocity
at $\bet (\bm n_{\bm p}) = 0$.
Here we used Eq.~(\ref{small})
in order to expand the square root.

\begin{figure}[t]
	\includegraphics[width=0.385\columnwidth]{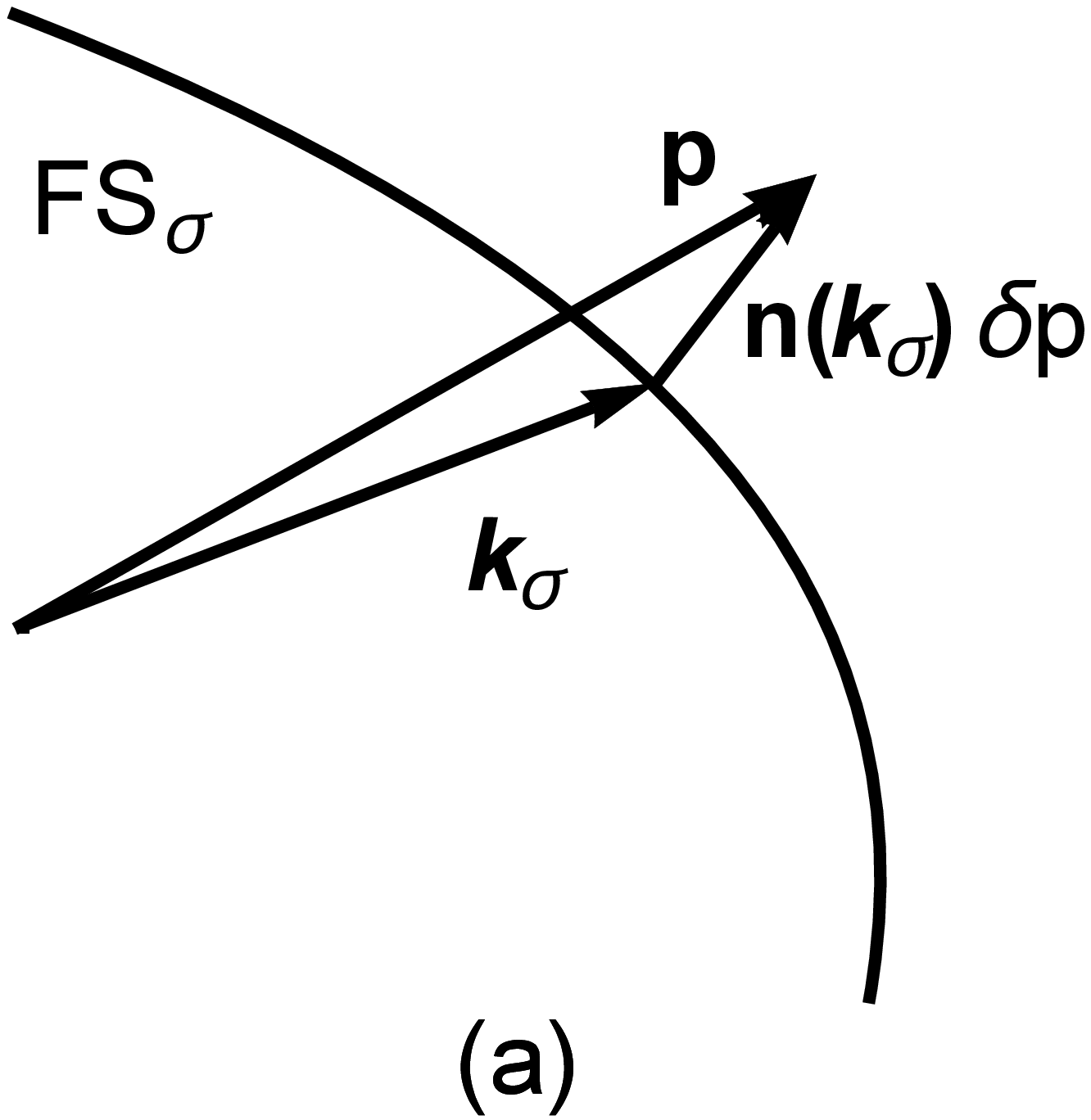}
	\hspace{3mm}
	\includegraphics[width=0.515\columnwidth]{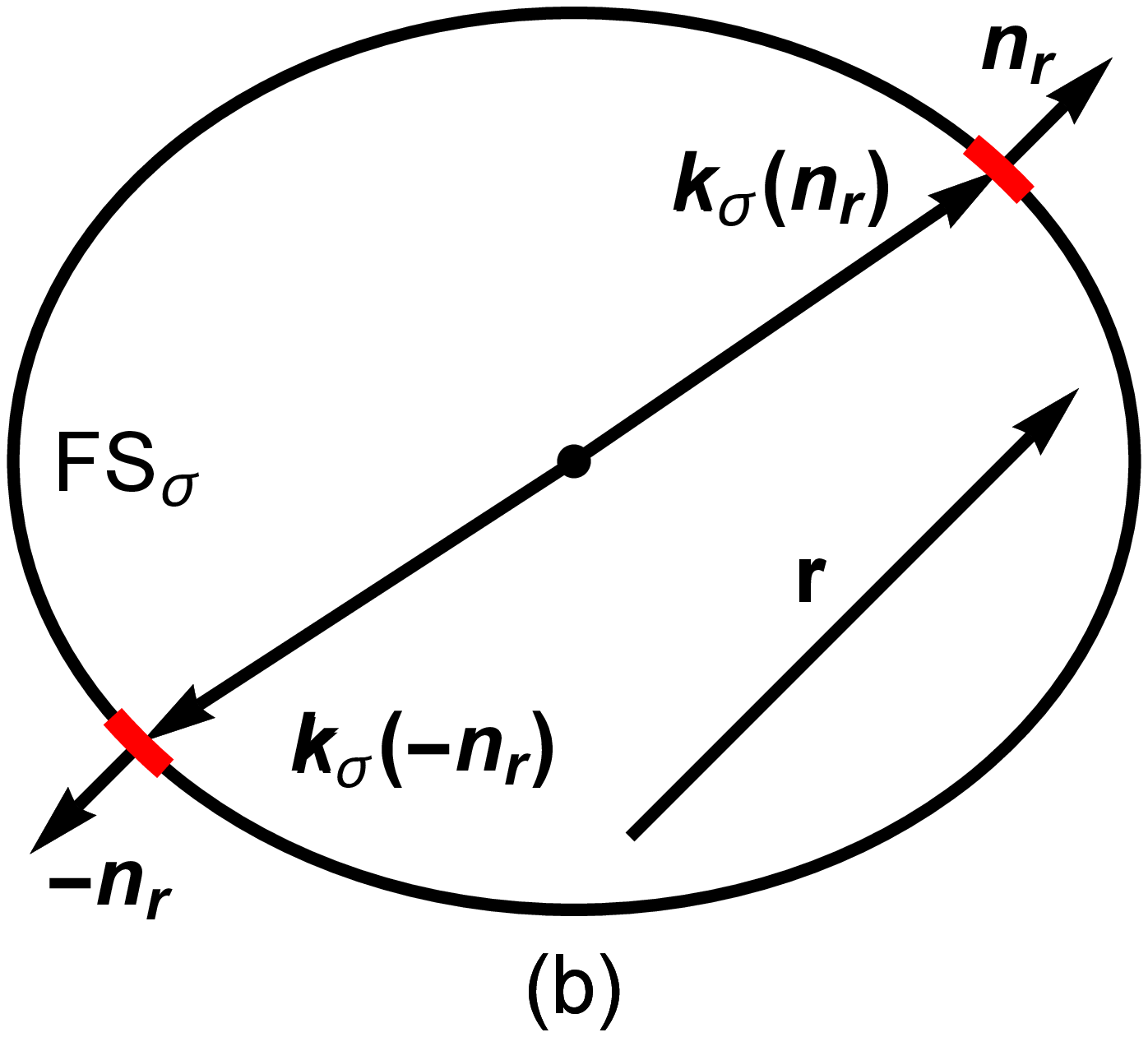}
	\caption{ 
		(a) Expansion of the momentum $\bm p$ close to the Fermi surface 
		$\mathcal{FS}_\sigma$:
		$\bm k_\sigma$ is the normal projection of $\bm p$ 
		on $\mathcal{FS}_\sigma$,
		$\bm n (\bm k_\sigma)$ is the outward normal 
		at $\bm k_\sigma \in \mathcal{FS}_\sigma$,
		$\delta p \ll k_F$.
		(b) Two points $\bm k_\sigma (\bm n_{\bm r})$ and $\bm k_\sigma (-\bm n_{\bm r})$
		on a nearly spherical Fermi surface 
		$\mathcal{FS}_\sigma$
		where the outward normals are equal to $\bm n_{\bm r}$
		and $- \bm n_{\bm r}$, respectively.
		The two red patches on $\mathcal{FS}_\sigma$ 
		correspond to the
		vicinities $U(\pm \bm n_{\bm r})$ of the points $\bm k_\sigma(\pm \bm n_{\bm r})$
		which 
		give the leading contribution to the $\tau \gg 1/E_F$ and $r \gg \lambda_F$
		asymptotics of the Green function, see Eq.~(\ref{G0}).
	}
	\label{fig:fsexp}
\end{figure}

\section{Asymptotics of the free electron Green function}

In the following it is most convenient to work with 
the electron Green function 
in the space-time representation.
In this paper we operate with the statistical (Matsubara) Green function $G_\sigma (\tau, \bm r)$,
where $\tau$ is the imaginary time,
$\bm r$ the $D$-dimensional coordinate vector, and
$\sigma = \pm 1$ the band index.
In this section we derive 
the asymptotics of the free 
electron Green function $G^{(0)}_\sigma (\tau, \bm r)$
at $\tau \gg 1/E_F$ and $r \gg \lambda_F$,
where $\lambda_F = 2 \pi / k_F$ 
is the Fermi wavelength.
Similar derivations can be found in Ref.~\cite{lounis}
in application to the Fermi surface imaging.

The asymptotics of 
$G^{(0)}_\sigma (\tau, \bm r)$
at $\tau \gg 1/E_F$ and $r \gg \lambda_F$
comes from the sector $(\omega, \bm p)$ close to the Fermi surface:
\begin{eqnarray}
&& \omega \ll E_F, \,\, \bm p = \bm k_\sigma + \bm n (\bm k_\sigma) \, \delta p, \,\, \delta p \ll k_F , \label{infrasec}
\end{eqnarray}
where $\bm k_\sigma \in \mathcal{FS}_\sigma$ 
is a point on the spin-split Fermi surface 
$\mathcal{FS}_\sigma$ with index $\sigma$,
$\bm n (\bm k_\sigma)$ is the outward normal at this point,
$\delta p > 0$ ($\delta p < 0$) corresponds to 
empty (occupied) states 
at zero temperature,
see Fig.~\ref{fig:fsexp}(a).
The free electron Green function $G_\sigma^{(0)}(i \omega, \bm p)$ 
is given by the quasiparticle pole:
\begin{eqnarray}
&& G_\sigma^{(0)} (i \omega, \bm p) \equiv  G_\sigma^{(0)}(i \omega, \delta p, \bm n) = \frac{|\sigma, \bm n\rangle \langle \sigma, \bm n|}{i\omega - v_\sigma (\bm n) \delta p} ,  \label{Gwp}
\end{eqnarray}
where we shortened $\bm n(\bm k_\sigma)$ to $\bm n$ here,
 $|\sigma, \bm n\rangle$
is the spinor given by Eq.~(\ref{psip}),
$v_\sigma (\bm n)$ is the Fermi velocity
at $\bm k_\sigma$.
Here we also linearized the dispersion
with respect to $\delta p$
because $\delta p \ll k_F$.
At the same time, the finite curvature
of the Fermi surface 
is important for the asymptotic form of $G^{(0)}_\sigma (\tau, \bm r)$.

The space-time representation of the free electron Green function is given by the Fourier transform:
\begin{eqnarray}
&& \hspace{-10pt}  G_\sigma^{(0)}(\tau, \bm r) = \int\limits_{-\infty}^\infty \frac{d\omega}{2 \pi} \int \frac{d \bm p}{(2 \pi)^D} e^{i (\bm p \cdot \bm r - \omega \tau)} G_\sigma^{(0)}(i \omega, \bm p) ,
\end{eqnarray}
where $G_\sigma^{(0)}(i \omega, \bm p)$
is given by Eq.~(\ref{Gwp})
at the infrared sector defined in Eq.~(\ref{infrasec}).
We integrate over the Matsubara frequencies because
here we consider the case of zero temperature $T = 0$.
The integral over $\omega$ is elementary:
\begin{eqnarray}
&& G_\sigma^{(0)}(\tau, \delta p, \bm n) = \int\limits_{-\infty}^\infty \frac{d\omega}{2 \pi} e^{-i \omega \tau} \frac{|\sigma, \bm n \rangle \langle \sigma, \bm n|}{i \omega - v_\sigma (\bm n) \delta p} \nonumber \\
&& \hspace{30pt} = - {\rm sgn}(\tau) \theta(\delta p \, \tau) e^{-v_\sigma (\bm n) \delta p \, \tau} |\sigma, \bm n \rangle \langle \sigma, \bm n| , \label{Gtk}
\end{eqnarray}
where ${\rm sgn}(\tau)$ returns the sign of $\tau$,
$\theta(z)$ is the Heaviside step function, 
i.e. $\theta(z) = 0$ ($\theta(z) = 1$)
if $z < 0$ ($z > 0$).

The integration over $\bm p$
is convenient to perform via thin
layers located at the distance $\delta p$ from the Fermi surface $\mathcal{FS}_\sigma$.
As $\delta p \ll k_F$,
the measure can be approximated as follows:
\begin{eqnarray}
&& d \bm p \approx d \bm k_\sigma \, d \delta p, \,\, \bm k_\sigma \in \mathcal{FS}_\sigma .
\end{eqnarray}
The momentum is expanded via Eq.~(\ref{infrasec}), see also Fig.~\ref{fig:fsexp}(a),
i.e. $\bm p = \bm k_\sigma + \bm n (\bm k_\sigma) \, \delta p$,
where the normal $\bm n (\bm k_\sigma)$
is taken at the point $\bm k_\sigma \in \mathcal{FS}_\sigma$.

Next, we apply the stationary phase method in order to evaluate the 
integral over $\bm k_\sigma \in \mathcal{FS}_\sigma$.
For this, we first find the stationary points 
where $\bm k_\sigma \cdot \bm r$ reaches its extrema.
This happens when $d \bm k_\sigma \cdot \bm r = 0$, where $d \bm k_\sigma$ is an arbitrary element 
of the tangent space attached to the Fermi surface $\mathcal{FS}_\sigma$
at the point $\bm k_\sigma \in \mathcal{FS}_\sigma$.
This condition is satisfied at such points 
$\bm k_\sigma \in \mathcal{FS}_\sigma$
at which the normals $\bm n(\bm k_\sigma)$
are collinear with the coordinate vector $\bm r$.
As the Fermi surfaces are nearly spherical,
see Eq.~(\ref{ksig}),
there are exactly two points $\bm k_\sigma (\pm \bm n_{\bm r}) \in \mathcal{FS}_\sigma$
where the outward normals 
are equal to $\pm \bm n_{\bm r}$, $\bm n_{\bm r} = \bm r/ r$ is the unit vector along $\bm r$, 
see Fig.~\ref{fig:fsexp}(b).
Thus, we find that the integral over $\bm p$
yields the sum of two integrals 
over small vicinities $U(\pm \bm n_{\bm r}) \subset \mathcal{FS}_\sigma$ of the points $\bm k_\sigma (\pm \bm n_{\bm r}) \in \mathcal{FS}_\sigma$,
see Fig.~\ref{fig:fsexp}(b):
\begin{widetext}
\begin{eqnarray}
&& G_\sigma^{(0)}(\tau, \bm r) \approx \int\limits_{\bm k_\sigma \in U(\bm n_{\bm r})} \frac{d \bm k_\sigma}{(2 \pi)^{D - 1}} e^{i (\bm k_\sigma - \bm k_\sigma (\bm n_{\bm r})) \cdot \bm r} \int\limits_{-\infty}^\infty \frac{d\delta p}{2 \pi} e^{i \delta p \, r} G_\sigma^{(0)}(\tau, \delta p, \bm n_{\bm r}) e^{i \bm k_\sigma (\bm n_{\bm r}) \cdot \bm r} \nonumber \\
&& + \int\limits_{\bm k_\sigma \in U(-\bm n_{\bm r})} \frac{d \bm k_\sigma}{(2 \pi)^{D - 1}} e^{i (\bm k_\sigma - \bm k_\sigma (-\bm n_{\bm r})) \cdot \bm r} \int\limits_{-\infty}^\infty \frac{d\delta p}{2 \pi} e^{-i \delta p \, r} G_\sigma^{(0)}(\tau, \delta p, -\bm n_{\bm r}) e^{i \bm k_\sigma (-\bm n_{\bm r}) \cdot \bm r} , \,\,\,\, \bm n_{\bm r} = \frac{\bm r}{r} . \label{Gsum}
\end{eqnarray}
\end{widetext}
Here $\bm k_\sigma (\pm \bm n_{\bm r})$ 
are the points on $\mathcal{FS}_\sigma$
where the outward normals are equal to $\pm \bm n_{\bm r}$,
see Fig.~\ref{fig:fsexp}(b).
The integration over $\delta p$ is extended to the interval $\delta p \in (-\infty, \infty)$
due to quick convergence on the scale $\delta p \sim 1/r \ll k_F$.
The integrals over $\bm k_\sigma$ 
in the vicinities $U(\pm \bm n_{\bm r}) \subset \mathcal{FS}_\sigma$ of
the points $\bm k_\sigma(\pm \bm n_{\bm r})$ 
are Gaussian and they are convergent due to the finite Gaussian curvature of nearly spherical Fermi surface $\mathcal{FS}_\sigma$
at the points $\bm k_\sigma (\pm \bm n_{\bm r})$,
see Appendix \ref{app} for more details . 
The integrals over $\delta p$ yield the one-dimensional Fourier transforms:
\begin{eqnarray}
&& G_\sigma^{(0)}(\tau, x, \bm n) = \int\limits_{-\infty}^\infty \frac{d\delta p}{2\pi} e^{i \delta p \, x} G_\sigma^{(0)}(\tau, \delta p, \bm n) \nonumber \\
&& = \frac{1}{2 \pi}\frac{|\sigma, \bm n\rangle \langle \sigma, \bm n|}{i x - v_\sigma (\bm n) \tau} , \label{Gxt}
\end{eqnarray}
where $\bm n$ here is an arbitrary unit vector and
$x \in (-\infty, \infty)$ an effective one-dimensional coordinate.
Keeping only the linear order in SO splitting,
we show in the Appendix \ref{app} that
\begin{eqnarray}
&& \bm k_\sigma (\pm \bm n_{\bm r}) \cdot \bm r \approx \pm k_\sigma (\pm \bm n_{\bm r}) r , \label{kdotr}
\end{eqnarray}
where $k_\sigma (\bm n)$ is given by Eq.~(\ref{ksig}) for arbitrary unit vector $\bm n$.
The Gaussian integrals over $U(\pm \bm n_{\bm r})$
are proportional to $1/r^{(D - 1)/2}$,
see Appendix \ref{app} for the details.
Substituting Eqs.~(\ref{Gxt}) and  (\ref{kdotr})
into Eq.~(\ref{Gsum})
and evaluating the Gaussian integrals over $U(\pm \bm n_{\bm r})$, 
we find the infrared long-range asymptotics of the free-electron Matsubara Green function:
\begin{eqnarray}
&& G^{(0)}_\sigma (\tau, \bm r) \approx  \left(\frac{1}{\lambda_F r}\right)^{\frac{D - 1}{2}}  \left[\frac{e^{i (k_\sigma (\bm n_{\bm r}) r - \vartheta)} }{2\pi} \frac{|\sigma, \bm n_{\bm r}\rangle \langle \sigma, \bm n_{\bm r}|}{ir - v_F \tau} \right. \nonumber \\
&& \left. - \frac{e^{-i (k_\sigma (-\bm n_{\bm r}) r - \vartheta)} }{2\pi} \frac{|\sigma, -\bm n_{\bm r}\rangle \langle \sigma, -\bm n_{\bm r}|}{ir + v_F \tau}\right] , \,\, \bm n_{\bm r} = \frac{\bm r}{r} , \label{G0} \\
&& \vartheta = \frac{\pi}{4} (D - 1) , \label{vartheta1}
\end{eqnarray}
where $k_\sigma (\bm n)$ is given by Eq.~(\ref{ksig})
for arbitrary unit vector $\bm n$,
$\lambda_F$ is the Fermi wavelength,
$v_F = k_F / m$ is the Fermi velocity 
at zero spin splitting,
and $\bm n_{\bm r} = \bm r/ r$
is the unit vector along $\bm r$.
Here we stress that Eq.~(\ref{G0})
is true only if the SO splitting is small compared to $E_F$, see Eq.~(\ref{small}).
We also neglected the weak dependence 
of the Fermi velocity on the 
spin splitting in the denominators in Eq.~(\ref{G0}) 
because it does not provide any non-analyticities.
We see from Eq.~(\ref{G0})
that the Green function 
contains the oscillatory factors that are sensitive to the spin splitting through $k_\sigma(\pm \bm n_{\bm r})$, see Eq.~(\ref{ksig}).
As we will see later,
these oscillatory factors are responsible
for the non-analytic terms in the thermodynamic potential $\Omega$.

In the Appendix \ref{app}  we generalize
this calculation to the case of a strongly interacting 
electron gas with a singularity (not necessarily a pole) at the Fermi surface
of arbitrary geometry.
The Appendix \ref{app} also contains details for nearly spherical Fermi surfaces.
The case of spherical Fermi surfaces 
is covered in Ref.~\cite{miserev}.

\section{Non-analyticities in \mbox{\boldmath{$\Omega$}}: limit of weak interaction}

In this section we calculate the non-analytic corrections
to the thermodynamic potential $\Omega$
in the limit of weak electron-electron interaction.
The calculation is performed within  second order perturbation theory
and it is valid in the paramagnetic Fermi liquid phase
far away from the FQPT.
However, this calculation is important 
because it allows us to identify the 
resonant scattering processes
close to the Fermi surface which are responsible for the non-analytic terms in $\Omega$.
In the following sections we treat these processes
non-perturbatively
and find that the non-analytic terms
are strongly enhanced close to the FQPT.
The results of this section extend existing theories 
\cite{belitz1,belitz2,chubukov1,maslov,zak1}
to the case of arbitrary spin splitting.
In particular, we show that arbitrary SO splitting 
is not able to gap out all soft fluctuation modes
and the non-analyticity in $\Omega$ with respect to the
magnetic field $\bm B$ survives,
in contrast to predictions of Ref.~\cite{kirk}.
In this section we are only after 
the non-analytic terms in $\Omega$,
all analytic corrections will be dropped.

\begin{figure}[t]
	\centering
	\includegraphics[width=0.45\columnwidth]{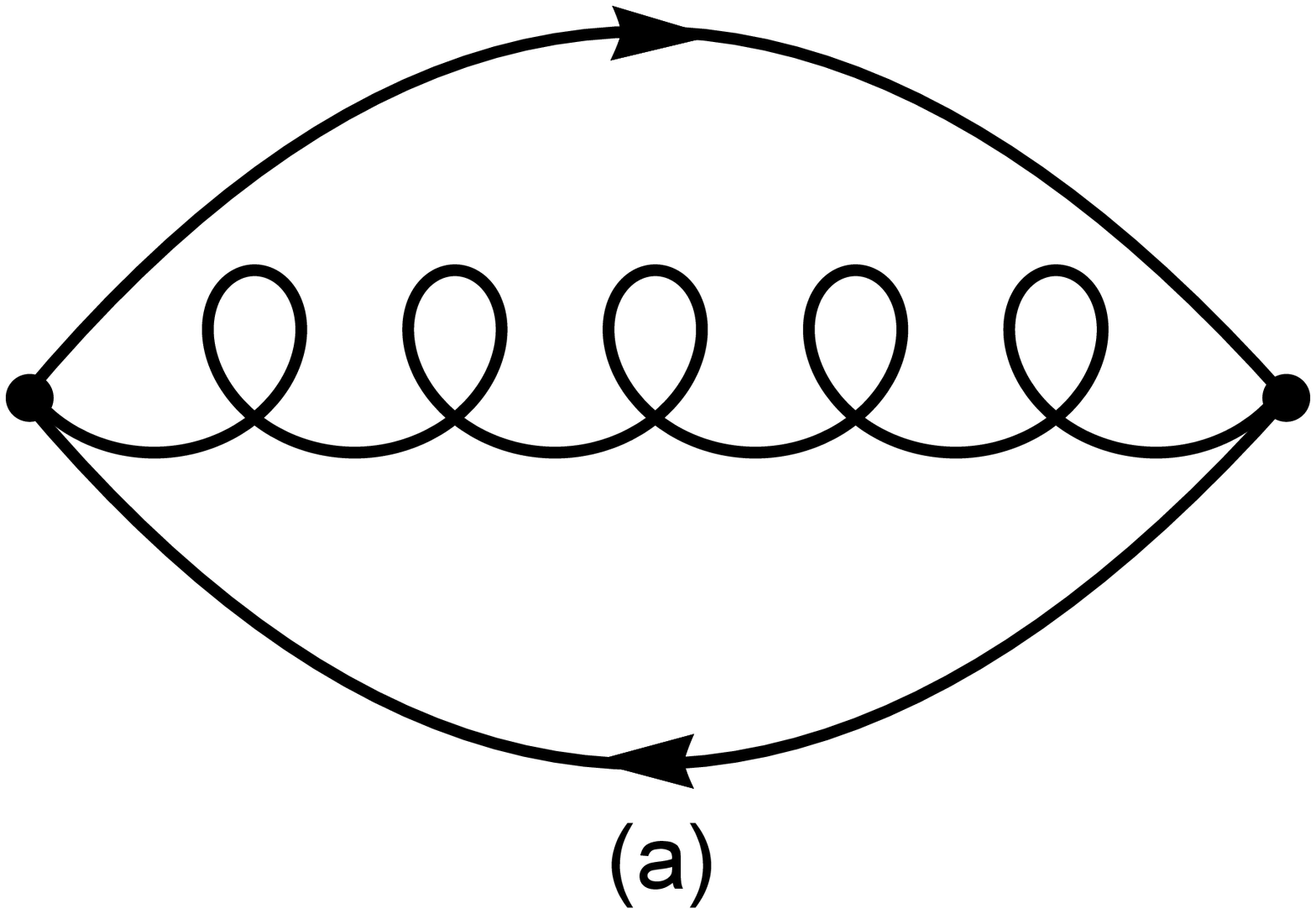}
	\hspace{3mm}
	\includegraphics[width=0.45\columnwidth]{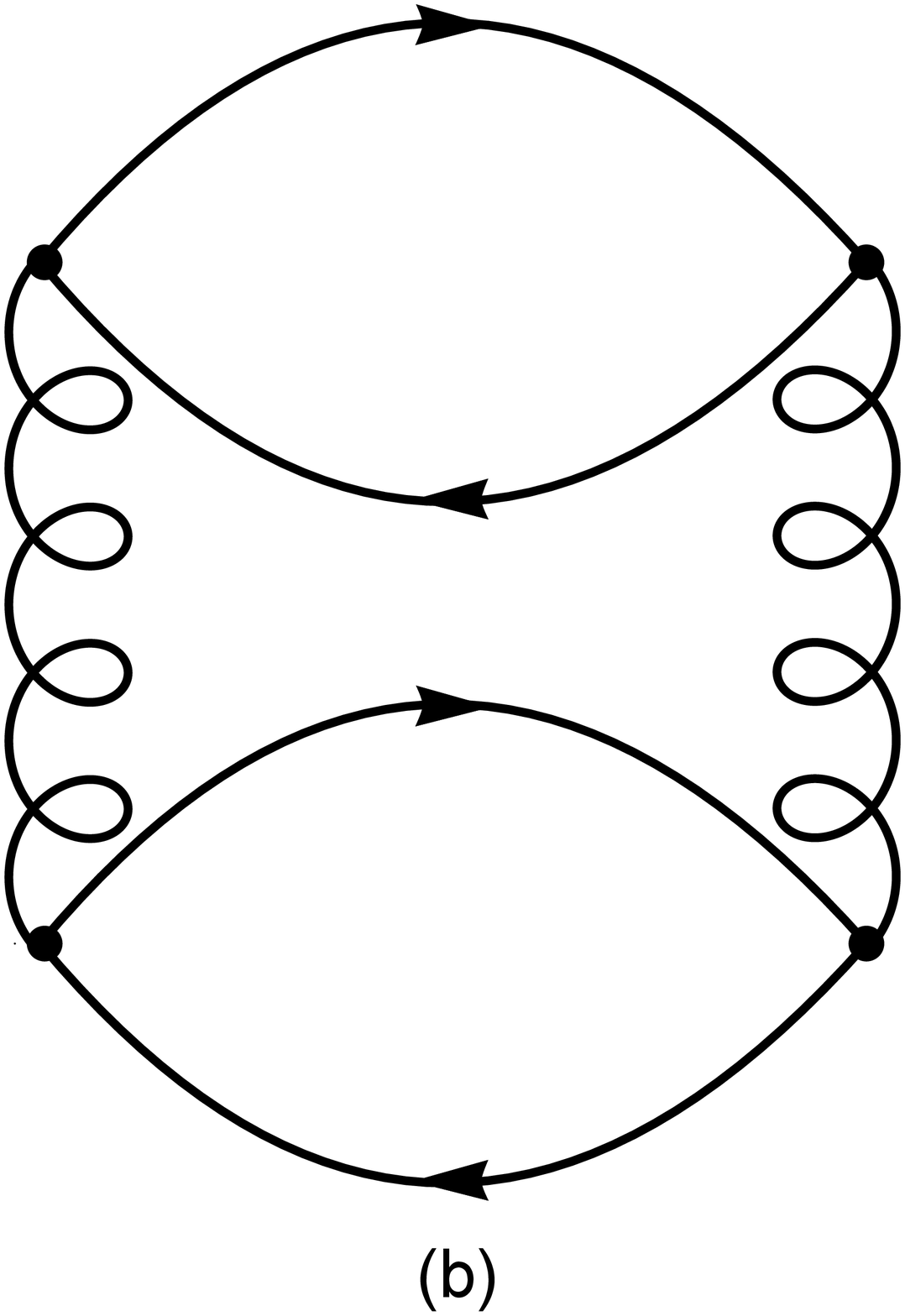}
	\caption{ 
		(a) First-order interaction correction to $\Omega$, see Eq.~(\ref{firstorder}).
		(b) Second-order interaction correction to $\Omega$ 
		contributing to the non-analyticity.
		Solid lines correspond to 
		the electron propagators $G_\sigma^{(0)}(\tau, \bm r)$, see Eq.~(\ref{G0});
		wiggly lines stand for  
		the Coulomb interaction, see Eq.~(\ref{coul}).
	}
	\label{fig:diag12}
\end{figure}

\subsection{First-order interaction correction to \mbox{\boldmath{$\Omega$}}}

Let us start from the first-order interaction correction to $\Omega$, see Fig.~\ref{fig:diag12}(a):
\begin{eqnarray}
&& \Omega^{(1)} = \frac{1}{2} \sum\limits_{\sigma, \sigma'} \int dz \, V_0(z) P_{\sigma\sigma'}(z) , \label{firstorder}\\
&& P_{\sigma\sigma'} (z) = -{\rm Tr} \left\{G^{(0)}_\sigma(z) G^{(0)}_{\sigma'} (-z) \right\}, \label{P} 
\end{eqnarray}
where $z = (\tau, \bm r)$,
${\rm Tr}$ stands for the spin trace, and
$P_{\sigma\sigma'}(z)$ is the particle-hole 
bubble. Here, 
$V_0(z)$ is the Coulomb interaction,
\begin{eqnarray}
&& V_0(\tau, \bm r) = \frac{e^2}{\epsilon r} \delta(\tau), \label{coul}
\end{eqnarray}
where $e$ is the elementary charge,
$\epsilon$ the dielectric constant,
and $\delta (\tau)$ is due to the instantaneous nature of the Coulomb interaction 
(the speed of light is much larger than the Fermi velocity).
Using the asymptotics of the 
Green function $G^{(0)}_\sigma(\tau, \bm r)$, see Eq.~(\ref{G0}),
we find the asymptotics of the 
particle-hole bubble:
\begin{eqnarray}
&& P_{\sigma \sigma'}(\tau, \bm r) = P^{L}_{\sigma \sigma'} (\tau, \bm r) + P^{K}_{\sigma \sigma'} (\tau, \bm r) , \label{P0} \\
&& P^{L}_{\sigma \sigma'} (\tau, \bm r) \approx \frac{\delta_{\sigma\sigma'}}{2 \pi^2} \left(\frac{1}{\lambda_F r}\right)^{D - 1}  \frac{v_F^2 \tau^2 - r^2}{(r^2 + v_F^2 \tau^2)^2} , \label{LD} \\
&& P^{K}_{\sigma\sigma'}(\tau, \bm r) \approx \frac{1}{4 \pi^2} \frac{1}{r^2 + v_F^2 \tau^2} \left(\frac{1}{\lambda_F r}\right)^{D - 1} \nonumber \\
&& \times \left[e^{-2 i \vartheta} e^{i r (k_\sigma(\bm n_{\bm r}) + k_{\sigma'}(-\bm n_{\bm r}))} \left|M_{\sigma \sigma'}(\bm n_{\bm r})\right|^2 \right. \nonumber \\
&& \left. + e^{2 i \vartheta} e^{-i r (k_\sigma(-\bm n_{\bm r}) + k_{\sigma'}(\bm n_{\bm r}))} \left|M_{\sigma \sigma'}(-\bm n_{\bm r})\right|^2 \right] , \label{PK}
\end{eqnarray}
where the matrix elements $M_{\sigma\sigma'}(\pm \bm n_{\bm r})$ are given by Eq.~(\ref{mss}).
Here $P^{L}_{\sigma\sigma'}(\tau, \bm r)$ is the Landau damping contribution to the 
particle-hole bubble
coming from the forward scattering.
It is clear that this contribution is insensitive to the spin splitting.
The second contribution, $P^{K}_{\sigma\sigma'}(\tau, \bm r)$,
is the Kohn anomaly coming from the backscattering with the momentum transfer close to $2 k_F$.
The Kohn anomaly is sensitive to the spin splitting through the oscillatory factors
containing the Fermi momenta $k_\sigma (\pm \bm n_{\bm r})$, see Eq.~(\ref{ksig}).

As only the Kohn anomaly is sensitive to the spin splitting $\bet(\bm n_{\bm p})$,
we can simplify Eq.~(\ref{firstorder}):
\begin{eqnarray}
&& \Omega^{(1)} = \frac{1}{2} \sum\limits_{\sigma, \sigma'} \int\limits_{S_{D - 1}} d \bm n_{\bm r} \int\limits_0^\infty dr \, r^{D - 1} \frac{e^2}{\epsilon r}   P^K_{\sigma \sigma'} (0, \bm r) , \label{fo2}
\end{eqnarray}
where $S_{D - 1}$ is the $(D - 1)$-dimensional unit sphere,
$d \bm r = r^{D - 1} \, dr \, d \bm n_{\bm r}$.
The integral over $r$ is divergent at small $r$ 
(the ultraviolet divergence)
because it takes the following form:
\begin{eqnarray}
&& \int\limits_0^\infty \frac{dr}{r^3} e^{i r \Delta} \to \infty , \label{div}
\end{eqnarray}
where $\Delta$ is either equal to
$\Delta = k_\sigma(\bm n_{\bm r}) + k_{\sigma'}(-\bm n_{\bm r})$
or to $\Delta = -k_\sigma(-\bm n_{\bm r}) - k_{\sigma'}(\bm n_{\bm r})$.
This divergence comes from 
the asymptotics of the particle-hole bubble, see Eq.~(\ref{PK}),
that is only valid at $r \gg \lambda_F$.
Therefore, the lower limit for $r$ in Eq.~(\ref{div})
is bounded by $r \sim \lambda_F$.
This divergence can also be cured 
via the analytical continuation to the Euler gamma function $\Gamma (x)$
using the following identity:
\begin{eqnarray}
&& \hspace{-12pt}  I_\alpha (\Delta) = \int\limits_0^\infty \frac{dr}{r^\alpha} e^{i r \Delta} = \frac{\pi |\Delta|^{\alpha - 1}}{\sin(\pi \alpha) \Gamma(\alpha)} e^{-i \frac{\pi}{2} (\alpha - 1) {\rm sgn} (\Delta)} . \label{int}
\end{eqnarray}
In our case $\alpha = 3$ 
and the integral is indeed divergent due to
$\sin(3 \pi) = 0$ in the denominator.
Therefore, we consider $\alpha = 3 + \delta$
and take the limit $\delta \to 0$:
\begin{eqnarray}
&& \int\limits_0^\infty \frac{dr}{r^3} e^{i r \Delta} = \frac{\Delta^2}{2} \left( \frac{1}{\delta} + \ln \left|\Delta\right| - i \frac{\pi}{2} \sgn \left(\Delta\right)\right) .
\end{eqnarray}
Now it is clear that the physical dimension
of $\Delta$ under the logarithm
has to be compensated by the ultraviolet scale
$p_0 \sim 2 k_F$
which is equivalent to cutting the lower limit in Eq.~(\ref{div}) at $r \sim \lambda_F$:
\begin{eqnarray}
&& \int\limits_{\sim\lambda_F}^\infty \frac{dr}{r^3} e^{i r \Delta} = \frac{\Delta^2}{2} \left(\ln \left|\frac{\Delta}{p_0}\right| - i \frac{\pi}{2} \sgn \left(\Delta\right)\right) .
\end{eqnarray}
Now, we come back to $\Omega^{(1)}$
where $\Delta$ is either equal to
$\Delta = k_\sigma(\bm n_{\bm r}) + k_{\sigma'}(-\bm n_{\bm r})$
or to $\Delta = -k_\sigma(-\bm n_{\bm r}) - k_{\sigma'}(\bm n_{\bm r})$,
so using Eq.~(\ref{ksig}) we find:
\begin{eqnarray}
&& |\Delta| \approx 2 k_F + \frac{\sigma \beta(\pm \bm n_{\bm r}) + \sigma' \beta(\mp \bm n_{\bm r})}{v_F} .
\end{eqnarray}
As the spin splitting is much smaller than 
the Fermi energy,
we can expand the logarithm
$\ln|\Delta/p_0|$
in the analytic Taylor series with respect to the 
spin splitting.
Hence, we see that $\Omega^{(1)}$
does not contain any non-analyticities 
for arbitrary spin splitting.

Here we have performed the calculations for the 
long-range Coulomb interaction Eq.~(\ref{coul}).
Finite electron density results in the Thomas-Fermi screening of the 
long range Coulomb tail on the scale of the screening length $r_0$.
The weak coupling limit that we consider in this section
corresponds to $r_0 \gg \lambda_F$.
However, the
integral over $r$ in $\Omega^{(1)}$
converges at $r \sim \lambda_F \ll r_0$, because $\Delta \approx 2 k_F$ here.
Therefore, we can indeed neglect the Thomas-Fermi screening in this section.

\begin{figure}[t]
	\centering
	\includegraphics[width=0.45\columnwidth]{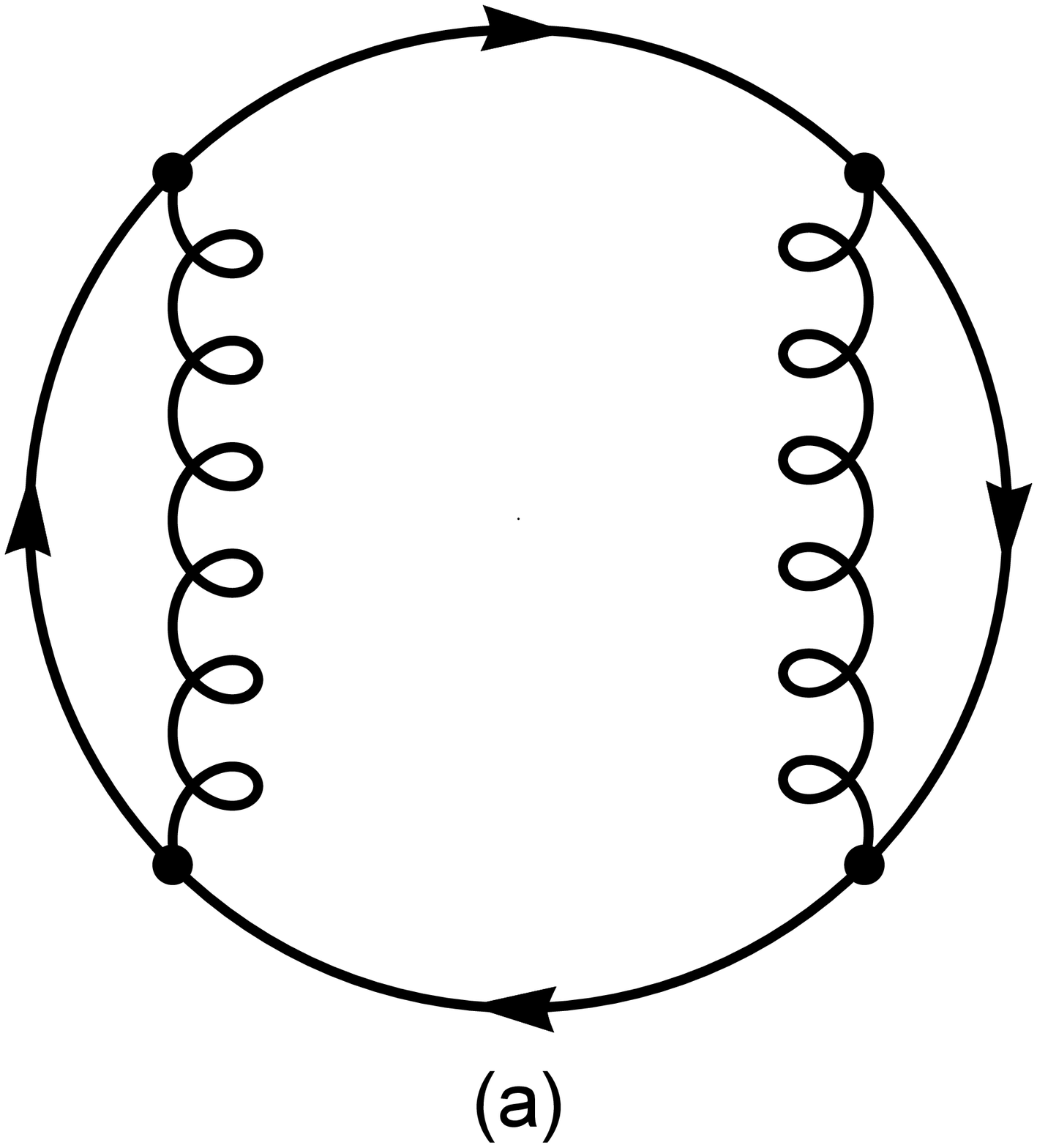}
	\hspace{3mm}
	\includegraphics[width=0.45\columnwidth]{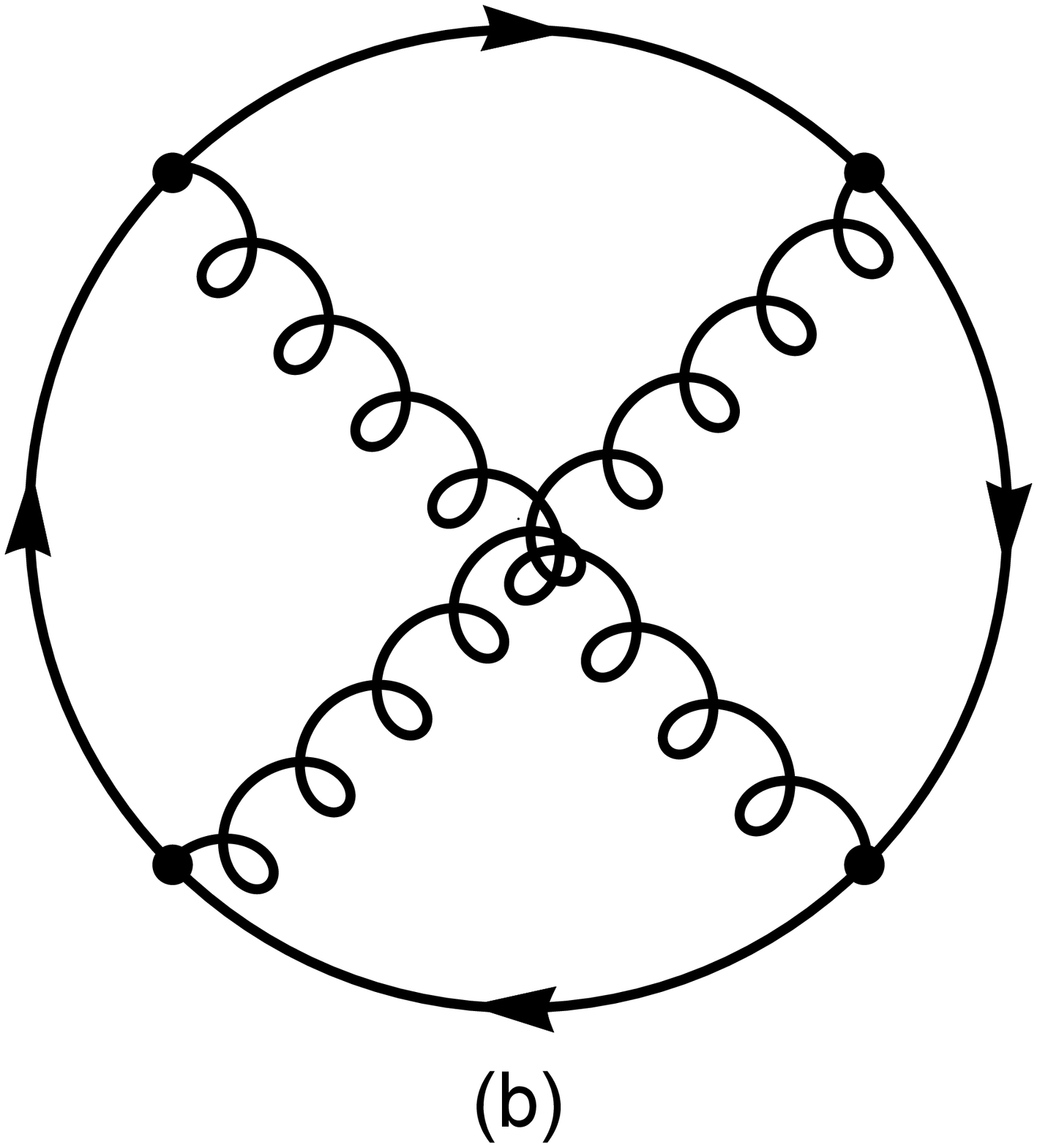}
	\caption{ 
		Other second-order diagrams that do not contribute to the non-analyticities in $\Omega$,
		see Eqs.~(\ref{omegaa}), (\ref{omegab}).
	}
	\label{fig:diagab}
\end{figure}

\subsection{Second-order interaction corrections to \mbox{\boldmath{$\Omega$}}}

We see from the calculation of $\Omega^{(1)}$
that the non-analytic terms 
may come from the oscillatory integrals 
like the one in Eq.~(\ref{int}).
However, we have to subtract the $2 k_F$
factor first, such that $\Delta$ 
in Eq.~(\ref{int}) becomes proportional to the spin splitting.
One way to achieve this
is to consider $\Omega^{(1)}$, see Eq.~(\ref{firstorder}), 
with the interaction $V(\tau, \bm r)$
which has oscillatory components $e^{\pm i 2 k_F r}$.
In fact, the electron-electron interaction 
acquires such components upon 
the dynamic screening by the particle-hole bubble.
One consequence of this is the Thomas-Fermi screening which we already discussed
and concluded that it is not important if the interaction is weak.
However, there is another much more important consequence of such dressing
that results in $2 k_F$ harmonics in the interaction
due to backscattering of electrons near the Fermi surface, the effect known as Friedel oscillations.
As we consider the correlations at large $r \sim v_F/\beta \gg \lambda_F$, where $\beta$ is a characteristic value of the spin splitting at the Fermi surface,
the interaction matrix elements at the momentum transfer $2 k_F$ are effectively local,
so we can use the effective contact interaction:
\begin{eqnarray}
&& V_{2 k_F}(z) = u \, \delta(\bm r) \delta (\tau) = u \, \delta (z), \label{contact} \\
&& u \approx V_0(2 k_F) = \frac{2 \pi^{\frac{D}{2}} \Gamma(D - 1)}{\Gamma\left(\frac{D}{2}\right)} 
\frac{e^2}{\epsilon (2 k_F)^{D - 1}} , \label{u}
\end{eqnarray}
where $\delta(\bm r)$ is the $D$-dimensional delta function,
$V_0(q)$ is the Fourier transform of the Coulomb 
interaction Eq.~(\ref{coul}).

If we dress the interaction line in Fig.~\ref{fig:diag12}(a)
by a single particle-hole bubble,
we get the second-order diagram for $\Omega$
shown in Fig.~\ref{fig:diag12}(b):
\begin{eqnarray}
&& \Omega^{(2)} = -\frac{1}{4} \sum\limits_{\sigma_i} \int dz_1 dz_2 dz_3 \, V_{2k_F}(z_2) V_{2 k_F}(z_3 - z_1) \nonumber \\
&& \hspace{70pt} \times P_{\sigma_1\sigma_2} (z_1)
P_{\sigma_3 \sigma_4}(z_3 - z_2) . \label{om0}
\end{eqnarray}
Using the contact approximation Eq.~(\ref{contact}),
we simplify $\Omega^{(2)}$ to the following expression:
\begin{eqnarray}
&& \Omega^{(2)} = - \frac{u^2}{4} \sum\limits_{\sigma_i} \int dz \, P_{\sigma_1 \sigma_2} (z) P_{\sigma_3 \sigma_4} (z) , \label{om2}
\end{eqnarray}
where $z = (\tau, \bm r)$.
From Eqs.~(\ref{P0})--(\ref{PK}) we find that only the product of Kohn anomalies
contains slowly oscillating terms
on the scale $v_F / \beta$, where $\beta$ stands for the characteristic spin splitting at the Fermi surface:
\begin{widetext}
\begin{eqnarray}
&& P_{\sigma_1 \sigma_2}(z) P_{\sigma_3 \sigma_4} (z) = \frac{e^{i r \Delta^{\sigma_1\sigma_2}_{\sigma_3\sigma_4} (\bm n_{\bm r})} \left|M_{\sigma_1 \sigma_2}(\bm n_{\bm r}) M_{\sigma_3 \sigma_4}(-\bm n_{\bm r})\right|^2 + 
e^{-i r \Delta^{\sigma_1\sigma_2}_{\sigma_3\sigma_4} (-\bm n_{\bm r})} \left|M_{\sigma_1 \sigma_2}(-\bm n_{\bm r}) M_{\sigma_3 \sigma_4}(\bm n_{\bm r})\right|^2}
{\left(2 \pi\right)^4 \left(\lambda_F r\right)^{2 (D - 1)} \left(r^2 + v_F^2 \tau^2\right)^2} + \dots , \label{PP} \\
&& \Delta^{\sigma_1 \sigma_2}_{\sigma_3\sigma_4} (\bm n_{\bm r}) = k_{\sigma_1} (\bm n_{\bm r}) + k_{\sigma_2} (-\bm n_{\bm r}) - k_{\sigma_3} (-\bm n_{\bm r}) -k_{\sigma_4} (\bm n_{\bm r}) \approx \frac{\left(\sigma_1 - \sigma_4\right) \beta (\bm n_{\bm r}) 
+ \left(\sigma_2 - \sigma_3\right) \beta (- \bm n_{\bm r})}{v_F} , \label{Delta}
\end{eqnarray}
\end{widetext}
where dots in Eq.~(\ref{PP}) stand for 
the rapidly oscillating terms on the scale of $2 k_F$ and $4 k_F$
and also the forward scattering contribution 
which does not contain any non-analytic dependence on the spin splitting.
We used Eq.~(\ref{ksig}) to express $\Delta^{\sigma_1\sigma_2}_{\sigma_3 \sigma_4} (\bm n_{\bm r})$
in terms of the spin splitting.

Then we substitute Eq.~(\ref{PP}) into Eq.~(\ref{om2})
and evaluate the integral over $z = (\tau, \bm r)$.
The integral over $\tau$ is elementary:
\begin{eqnarray}
&& \int\limits_{-\infty}^\infty \frac{d\tau}{\left(r^2 + v_F^2 \tau^2\right)^2} = \frac{\pi}{2 v_F r^3} . \label{tint}
\end{eqnarray}
The integral over $r$ can be represented using
the integral $I_\alpha (\Delta)$ defined in Eq.~(\ref{int}):
\begin{eqnarray}
&& \hspace{-10pt} \Omega^{(2)} = - \frac{u^2}{2^6 \pi^3 v_F \lambda_F^{2(D - 1)}} \sum\limits_{\sigma_i} \int\limits_{S_{D - 1}} d \bm n_{\bm r} \nonumber \\
&& \hspace{-10pt} \left|M_{\sigma_1 \sigma_2} (\bm n_{\bm r}) M_{\sigma_3 \sigma_4} (-\bm n_{\bm r})\right|^2 {\rm Re} \left(I_{D + 2} \left(\Delta^{\sigma_1\sigma_2}_{\sigma_3\sigma_4} (\bm n_{\bm r})\right)\right) , \label{om2dn}
\end{eqnarray}
where ${\rm Re}$ stands for the real part.
Here we used that $I_\alpha (-x) = I^*_\alpha (x)$,
where the star corresponds to the complex conjugation.

At this point it is convenient to introduce the 
dimensionless interaction parameter $g$:
\begin{eqnarray}
&& g = u N_F = \frac{u m k_F^{D - 2}}{2^{D - 1} \pi^{\frac{D}{2}} \Gamma\left(\frac{D}{2}\right)} , \label{coup}
\end{eqnarray}
where $N_F$ is the density of states per band at the Fermi level.
Substituting Eq.~(\ref{u}) into Eq.~(\ref{coup}),
we find an estimate for the dimensionless coupling constant $g$:
\begin{eqnarray}
&& g \approx \frac{\Gamma(D - 1)}{2^{2 D - 3} \Gamma^2 \left(\frac{D}{2}\right)} \frac{1}{k_F a_B}, \,\, a_B = \frac{\epsilon}{m e^2} , \label{coup2}
\end{eqnarray}
where $a_B$ is the effective Bohr radius.
The weak coupling regime corresponds to high densities such that $k_F a_B \gg 1$
or $g \ll 1$.

Then Eq.~(\ref{om2dn}) can be represented in the following form:
\begin{eqnarray}
&& \Omega^{(2)} = L_D \frac{v_F^{D + 1}}{2^{D + 2}} \sum\limits_{\sigma_i}
\int\limits_{S_{D-1}} d \bm n_{\bm r} \, 
\nonumber \\
&& \hspace{10pt} \times
\left|M_{\sigma_1\sigma_2}(\bm n_{\bm r}) M_{\sigma_3 \sigma_4}(-\bm n_{\bm r})\right|^2
 \left|\Delta^{\sigma_1 \sigma_2}_{\sigma_3\sigma_4}(\bm n_{\bm r})\right|^{D + 1} , \label{om2fin} \\
&& L_D = \frac{g^2}{32} \left(\frac{2}{\pi v_F}\right)^D \frac{\Gamma^2\left(\frac{D}{2}\right)}{\Gamma\left(D + 2\right)} \frac{1}{\cos \left(\pi \frac{D}{2}\right)} . \label{ld} 
\end{eqnarray}
We perform the summation over the band indexes $\sigma_i$ explicitly:
\begin{widetext}
	\begin{eqnarray}
	&& \Omega^{(2)} = L_D \int\limits_{S_{D - 1}} d \bm n \left[\left|M_{+-}(\bm n) M_{+-}(-\bm n)\right|^2 \left|\beta(\bm n) - \beta(-\bm n)\right|^{D + 1} + 
	\left|M_{++}(\bm n) M_{--}(\bm n)\right|^2 \left|\beta(\bm n) + \beta(- \bm n)\right|^{D + 1} \right. \nonumber \\
	&& \hspace{140pt} \left. + 
	2 \left(\left|M_{++}(\bm n) M_{-+}(\bm n)\right|^2 
	+ \left|M_{--}(\bm n) M_{+-}(\bm n)\right|^2\right) \left|\beta(\bm n)\right|^{D + 1}\right] . \label{omD}
	\end{eqnarray}
\end{widetext}
Here we dropped the index $\bm r$ in $\bm n_{\bm r}$,
such that $\bm n$ can be also interpreted as the 
unit vector $\bm n_{\bm p} = \bm p/p$, $p \approx k_F$,
in the momentum space.
This interpretation makes sense 
because the asymptotics of the Green function, see Eq.~(\ref{G0}),
comes from small vicinities of two points on the 
Fermi surface whose outward normals are collinear with $\bm r$.
So, $\bm r$ and $\bm p$ are in a way pinned to each other.

Finally, we have to check that 
the second-order diagrams in Fig.~\ref{fig:diagab}(a),(b)
do not contribute to the non-analytic terms in $\Omega$:
\begin{eqnarray}
&& \Omega_a = \frac{u^2}{2} \sum\limits_{\sigma_i} \int dz \, \nonumber \\
&& \hspace{30pt} {\rm Tr} \left\{G_{\sigma_1}^{(0)} (0) G_{\sigma_2}^{(0)} (z) G_{\sigma_3}^{(0)} (0) G_{\sigma_4}^{(0)} (-z) \right\} , \label{omegaa}\\
&& \Omega_b = \frac{u^2}{4} \sum\limits_{\sigma_i} \int dz \, \nonumber \\
&& \hspace{30pt} {\rm Tr} \left\{G_{\sigma_1}^{(0)} (z) G_{\sigma_2}^{(0)} (-z) G_{\sigma_3}^{(0)} (z) G_{\sigma_4}^{(0)} (-z) \right\} . \label{omegab}
\end{eqnarray}
Here $G_\sigma^{(0)} (0) = G_\sigma^{(0)} (\tau = -0, \bm r = 0)$ 
due to the ordering of the field operators within the interaction Hamiltonian:
\begin{eqnarray}
&& G_\sigma^{(0)} (0) = \int \frac{d \bm p}{(2 \pi)^D} \theta (k_\sigma (\bm n_{\bm p}) - p) |\sigma, \bm n_{\bm p} \rangle \langle \sigma, \bm n_{\bm p}| , 
\end{eqnarray}
where $|\sigma, \bm n_{\bm p}\rangle$
are the eigenvectors of the single-particle Hamiltonian, see Eq.~(\ref{psip}),
and $k_\sigma (\bm n_{\bm p})$ is given by Eq.~(\ref{ksig}).

The diagram $\Omega_a$ has a single particle-hole bubble in it
due to the Green functions $G_{\sigma_2}^{(0)}(z)$ and $G_{\sigma_4}^{(0)} (-z)$,
see Eq.~(\ref{omegaa}).
The product of these Green functions 
contains weakly oscillating terms
and $\approx 2k_F$ harmonics.
As in the case of $\Omega^{(1)}$,
the $2 k_F$ harmonics do not produce any non-analyticities.
The weakly oscillating terms originate from the 
Landau damping part of the particle-hole bubble
but these terms vanish due to the integral over $\tau$:
\begin{eqnarray}
&& \int\limits_{-\infty}^\infty \frac{d \tau}{(v_F \tau \pm i r)^2} = 0 .
\end{eqnarray}

The diagram $\Omega_b$ is more complicated.
Let us consider two matrix products $G_{\sigma_1}^{(0)}(z) G_{\sigma_2}^{(0)} (-z)$
and $G_{\sigma_3}^{(0)}(z) G_{\sigma_4}^{(0)} (-z)$,
spin traces are not taken here.
As usual, we are after the slowly oscillating terms in Eq.~(\ref{omegab}).
One possibility for this is
the product of the forward scattering contributions coming from
$G_{\sigma_1}^{(0)}(z) G_{\sigma_2}^{(0)} (-z)$
and $G_{\sigma_3}^{(0)}(z) G_{\sigma_4}^{(0)} (-z)$.
From Eq.~(\ref{G0})
it is clear that the forward scattering contributions
are non-zero only if $\sigma_1 = \sigma_2$
and $\sigma_3 = \sigma_4$,
matrix products of corresponding projectors vanish otherwise.
However, in this case the oscillating factors are canceled exactly
and thus, this contribution is analytic.
Another way to obtain slowly oscillating terms in Eq.~(\ref{omegab})
is the product of Kohn anomalies 
contained in $G_{\sigma_1}^{(0)}(z) G_{\sigma_2}^{(0)} (-z)$
and $G_{\sigma_3}^{(0)}(z) G_{\sigma_4}^{(0)} (-z)$.
In this case, we have to look at the spin trace in Eq.~(\ref{omegab}) 
which is non-zero only if $\sigma_1 = \sigma_4$
and $\sigma_2 = \sigma_3$.
This condition becomes obvious 
if we notice that the product of Kohn anomalies 
of $G_{\sigma_1}^{(0)}(z) G_{\sigma_2}^{(0)} (-z)$
and $G_{\sigma_3}^{(0)}(z) G_{\sigma_4}^{(0)} (-z)$
is actually equivalent to the product of the 
forward scattering contributions of 
$G_{\sigma_2}^{(0)}(-z) G_{\sigma_3}^{(0)} (z)$
and $G_{\sigma_4}^{(0)}(-z) G_{\sigma_1}^{(0)}(z)$
which is analytic for the reasons we discussed above.

Hence, only the diagram in Fig.~\ref{fig:diag12}(b)
contains  non-analytic terms
and, therefore, Eq.~(\ref{omD}) describes 
the non-analytic corrections to $\Omega$ 
due to arbitrary spin splitting $\bet (\bm p)$
within  second-order perturbation theory.

Even though Eq.~(\ref{omD}) is true in arbitrary number $D$ of spatial dimensions,
we give explicit expressions for $D = 2$ and $D = 3$.
For 2DEG the coefficient $L_2$
is negative, see Eq.~(\ref{ld}) for $D = 2$:
\begin{eqnarray}
&& L_2 = -\frac{g^2}{48 \pi^2 v_F^2} . \label{l2}
\end{eqnarray}
The integral over $d\bm n$ can be parametrized 
by a single angle $\phi \in (0, 2\pi]$,
so the non-analytic correction Eq.~(\ref{omD}) 
for 2DEG then reads:
\begin{widetext}
	\begin{eqnarray}
	&& \Omega^{(2)} = - \frac{g^2}{24 \pi v_F^2} \int\limits_0^{2\pi} \frac{d \phi}{2\pi} \left[ \left|M_{+-}(\phi) M_{-+}(\phi)\right|^2  \left|\beta(\phi) - \beta(\phi + \pi)\right|^3
	+
	\left|M_{++}(\phi) M_{--}(\phi)\right|^2  \left|\beta(\phi) + \beta(\phi + \pi)\right|^3 \right. \nonumber \\
	&& \hspace{95pt} \left. + 2 \left(\left|M_{++}(\phi) M_{-+}(\phi)\right|^2 + \left|M_{--}(\phi) M_{+-}(\phi)\right|^2\right) \left|\beta(\phi)\right|^3 \right] , \,\,\,\, D = 2 . \label{om2deg}
	\end{eqnarray}
\end{widetext}
Our result Eq.~(\ref{om2deg})
agrees
with  previous studies \cite{zak1,zak2,maslov}
and extends them to the case of arbitrary 
spin splitting.
Equation~(\ref{om2deg})
together with Eq.~(\ref{mss})
for the matrix elements
allows one to find the non-analytic terms in $\Omega$
directly from the spin splitting 
$\bet(\bm n)$.

The case of $D = 3$ is marginal because 
the non-analytic terms in Eq.~(\ref{omD})
are proportional to the fourth power of the 
spin splitting.
The non-analyticity itself 
comes from the divergence of the $L_D$ prefactor
at $D = 3$, see Eq.~(\ref{ld}),
which results in an additional logarithm.
This is best seen from the dimensional regularization:
\begin{eqnarray}
&& D = 3 - \delta, \,\, \delta \to +0.
\end{eqnarray}
The dimension $D$ enters Eq.~(\ref{omD})
in the following form:
\begin{eqnarray}
&& L_D \Delta^{D + 1} = -\frac{g^2 \Delta^4}{192 \pi^3 v_F^3} \left( \frac{1}{\delta} - \ln \Delta \right) + \mathcal{O}(\delta) ,
\end{eqnarray}
where $\Delta$ takes one of the following
values: $\Delta = |\beta(\bm n) \pm \beta(-\bm n)|$
or
$\Delta = |\beta(\bm n)|$.
Here we expanded the expression at $\delta \to + 0$.
The divergent $1/\delta$ contribution is actually analytic
and can be represented by $\ln \Lambda$ 
factor, $\Lambda \sim E_F$,
which compensates the physical dimension of $\Delta$:
\begin{eqnarray}
&& L_D \Delta^{D + 1} \to \frac{g^2}{48 \pi^2 v_F^3} \frac{\Delta^4}{4 \pi} \ln \left|\frac{\Delta}{\Lambda}\right| , \,\, \Lambda \sim E_F . \label{reg}
\end{eqnarray}
Using the regularization Eq.~(\ref{reg}),
we find the non-analytic correction to the spin-split 3DEG:
\begin{widetext}
	\begin{eqnarray}
	&& \Omega^{(2)} = \frac{g^2}{48 \pi^2 v_F^3} \int\limits_{S_2} \frac{d \bm n}{4 \pi} \left[\left|M_{+-}(\bm n) M_{-+}(\bm n)\right|^2 \left|\beta(\bm n) - \beta(-\bm n)\right|^4 \ln\left|\frac{\beta(\bm n) - \beta (-\bm n)}{\Lambda}\right| \right.
	\nonumber \\
	&& +
	\left|M_{++}(\bm n) M_{--}(\bm n)\right|^2 \left|\beta(\bm n) + \beta(-\bm n)\right|^4 \ln\left|\frac{\beta(\bm n) + \beta (-\bm n)}{\Lambda}\right| 
	\nonumber \\
	&& \left.
	+ 2 \left(\left|M_{++}(\bm n) M_{-+}(\bm n)\right|^2 + \left|M_{--}(\bm n) M_{+-}(\bm n)\right|^2\right) \left|\beta(\bm n)\right|^4 \ln\left|\frac{\beta (\bm n)}{\Lambda}\right| \right] , \,\,\,\, \Lambda \sim E_F , \,\,\,\, D = 3 . \label{om3deg}
	\end{eqnarray}
\end{widetext}
Here, integration over the unit sphere $S_2$
means $d \bm n = \sin \phi_1 \, d \phi_1 d\phi_2$,
$\phi_1 \in [0, \pi]$, $\phi_2 \in (0, 2\pi]$.
The non-analytic correction is negatively defined for arbitrary spin splitting
due to the logarithms.
In particular,
if $\bet(\bm n) = \bm B$,
we get the well-known result, see Ref.~\cite{maslov}:
\begin{eqnarray}
&& \Omega^{(2)} = \frac{g^2 B^4}{3 \pi^2 v_F^3}  \ln\left|\frac{2 B}{\Lambda}\right| , \,\, \Lambda \sim E_F . \label{B3D}
\end{eqnarray}

\subsection{Large SO splitting and small magnetic field}

Here, we consider the important special case 
of arbitrary SO splitting and small magnetic field:
\begin{eqnarray}
&& \bet (\bm n_{\bm p}) = \bet_{SO} (\bm n_{\bm p}) + \bm B , \,\, B \ll \beta_{SO}  ,  \label{bet}
\end{eqnarray}
where $\bm n_{\bm p} = \bm p / p$, $p \approx k_F$, 
$\beta_{SO}$ is a characteristic 
value of the SO splitting at the Fermi surface.
As any SO splitting respects time reversal symmetry,
it has to be an odd vector function of $\bm n_{\bm p}$:
\begin{eqnarray}
&& \bet_{SO}(-\bm n_{\bm p}) = - \bet_{SO} (\bm n_{\bm p}) . \label{tr}
\end{eqnarray}
As we consider $B \ll \beta_{SO}$, then
we can expand $\beta (\bm n)$ with respect to $\bm B$:
\begin{eqnarray}
&& \beta (\bm n) \approx \beta_{SO} (\bm n) + \frac{\bet_{SO}(\bm n) \cdot \bm B}{\beta_{SO} (\bm n)} , 
\end{eqnarray}
where $\beta_{SO}(\bm n) = |\bet_{SO}(\bm n)|$.
Together with the symmetry condition Eq.~(\ref{tr}), we 
conclude that only the very first term in Eq.~(\ref{omD}) 
contributes to the non-analyticity  with respect to $\bm B$
due to the following identity:
\begin{eqnarray}
&& \beta(\bm n) - \beta(-\bm n) \approx 2 \frac{\bet_{SO}(\bm n) \cdot \bm B}{\beta_{SO} (\bm n)} . \label{db}
\end{eqnarray}
As we only consider the leading non-analyticity,
we calculate the matrix elements 
at $\bm B = 0$:
\begin{eqnarray}
&& M_{\sigma \sigma} (\bm n) = 0, \,\, M_{\sigma -\sigma} (\bm n) = -1 . \label{mso}
\end{eqnarray}
Substituting Eqs.~(\ref{db}), (\ref{mso})
in Eq.~(\ref{omD}), we find
the non-analytic in $\bm B$ correction to $\Omega$ in case of arbitrary SO splitting:
\begin{eqnarray}
&& \delta\Omega (\bm B) = L_D \int\limits_{S_{D - 1}} d\bm n \, \left|2\frac{\bet_{SO} (\bm n) \cdot \bm B}{\beta_{SO}(\bm n)}\right|^{D + 1} , \label{sogen}
\end{eqnarray}
where $\delta \Omega (\bm B)$ indicates that 
only the non-analytic terms with respect to $\bm B$ are included.
Thus, we see that the non-analyticity in
magnetic field 
$\bm B$ cannot be eliminated even by arbitrary SO splitting,
in contrast to the predictions of Ref.~\cite{kirk}.

\begin{figure}[t]
	\centering
	\includegraphics[width=0.45\columnwidth]{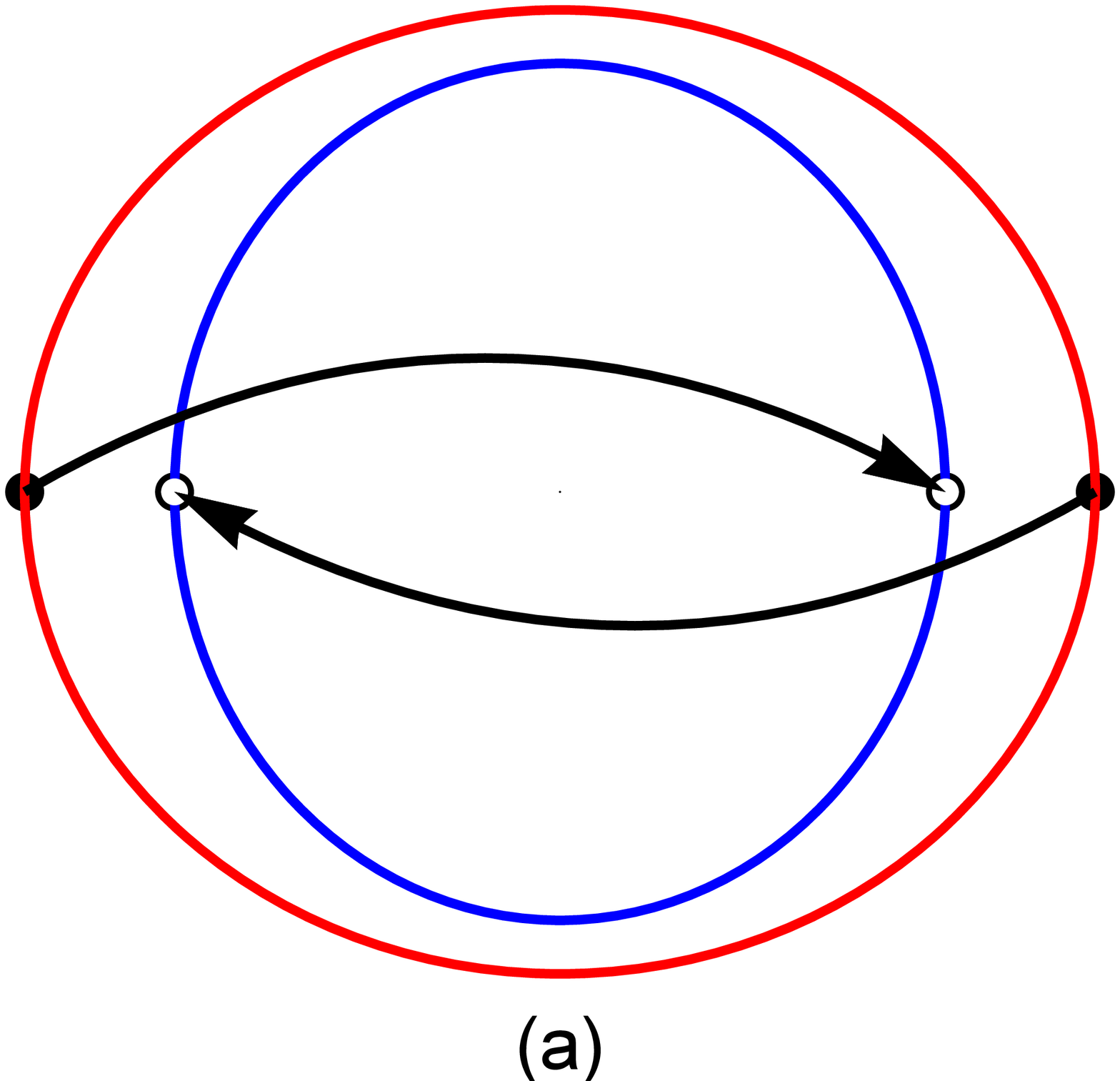}
	\hspace{3mm}
	\includegraphics[width=0.45\columnwidth]{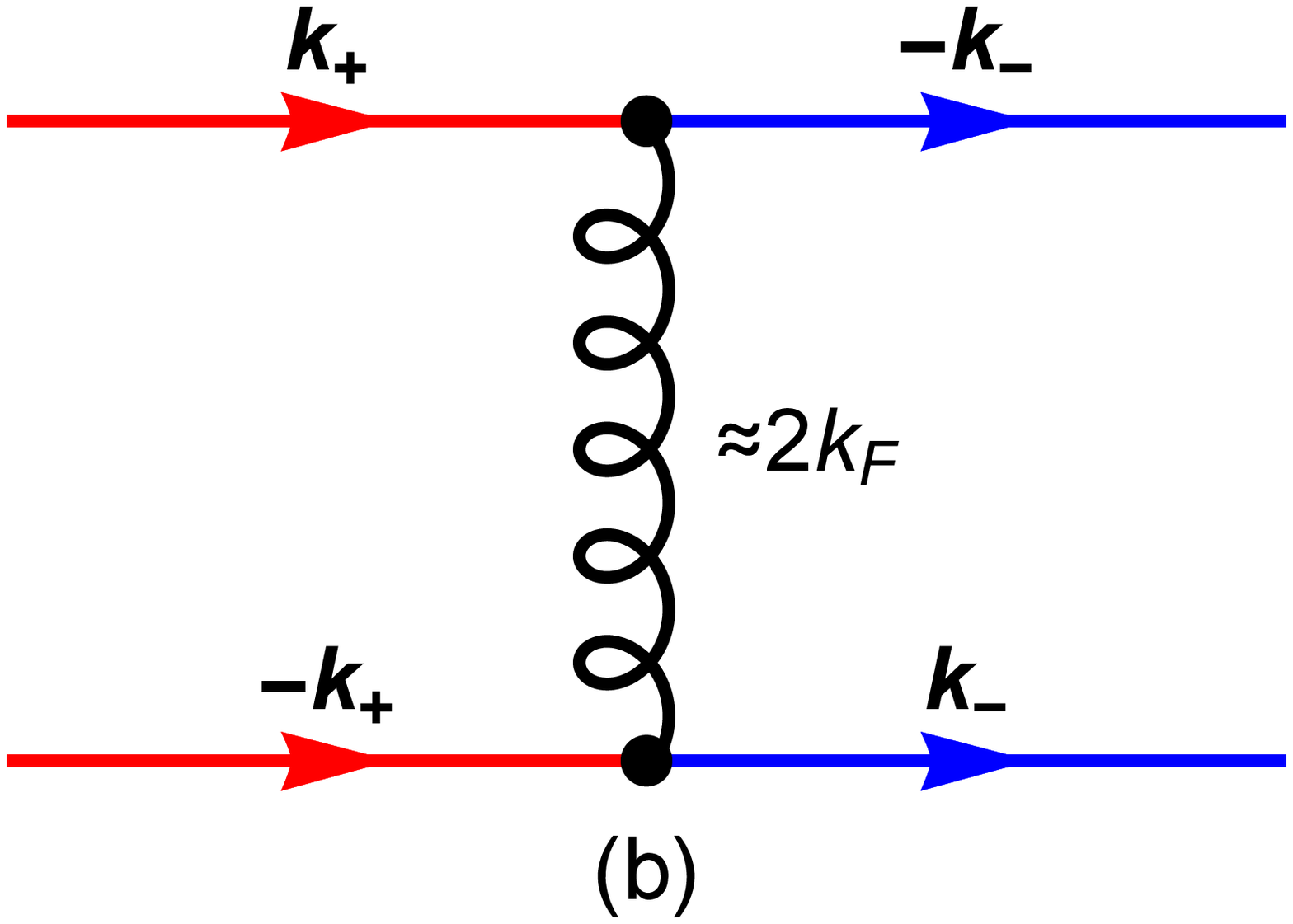}
	\caption{ 
		(a) Fermi surfaces at arbitrary SO splitting, red (blue) 
		color corresponds to $\sigma = +1$ ($\sigma = -1$). 
		The arrows show the resonant scattering processes.
		(b) The interaction matrix element 
		corresponding to the 
		resonant scattering processes
		at finite SO splitting.
		Here a pair of electrons
		with the band index $\sigma = +1$ and opposite momenta $\pm \bm k_+$
		scatter into a pair 
		with momenta $\pm \bm k_{-}$
		that are collinear with $\pm \bm k_+$.
		These processes are resonant due to the time reversal symmetry, see Eq.~(\ref{tr}). 
		The collinearity of $\bm k_+$ and $\bm k_-$ is due to the {\it local nesting} discussed in the main text after Eq.~(\ref{sogen}).
		These processes are responsible for the non-analyticity in $\Omega$ with respect to 
		small magnetic field $\bm B$, see Eq.~(\ref{sogen}).
	}
	\label{fig:SOres}
\end{figure}

The elementary processes that are responsible for the non-analyticity given by Eq.~(\ref{sogen})
are shown in Fig.~\ref{fig:SOres}.
These processes describe the 
resonant scattering of a pair of electrons with
the band index $\sigma$
and opposite momenta $\pm \bm k_\sigma$
into a pair of electrons in the other band with index $-\sigma$ 
and momenta $\pm \bm k_{-\sigma}$ 
that are collinear with momenta of initial electrons $\pm \bm k_\sigma$.
The momentum transfer in such a scattering processes is close to $2 k_F$, see Fig.~\ref{fig:SOres}(b).
The scattering with small momentum transfer is forbidden 
due to the orthogonality condition Eq.~(\ref{orthogonality}).
The considered processes are resonant due to the 
time reversal symmetry, see Eq.~(\ref{tr}).
The collinearity condition comes from 
the {\it local nesting}
when the momentum transfer between the resonantly
scattering states 
also matches small vicinities 
around these states.
This matching is satisfied when the outward normals
in the scattering states are collinear
such that the mismatch comes only from different curvatures of the Fermi surface
in the considered points.
The local nesting strongly enhances
corresponding scattering processes
because not only the considered states are in  resonance but also small vicinities of states around them.
For example, the Kohn anomaly in the particle-hole bubble
is a result of such a local nesting 
for the states scattering with the $2 k_F$ momentum transfer.
The perfect local nesting 
corresponds to the Landau damping of the particle-hole excitations
with energy and momentum around zero,
in this case the scattered region in the particle-hole bubble is mapped onto itself.

It is also instructive to 
write down Eq.~(\ref{sogen}) 
for 2DEG 
and 3DEG explicitly:
\begin{eqnarray}
&& \delta\Omega(\bm B) = - \frac{g^2 |B|^3}{3 \pi v_F^2} \int\limits_0^{2\pi} \frac{d\phi}{2 \pi} \left|\frac{\bet_{SO} (\phi) \cdot \bm b}{\beta_{SO}(\phi)}\right|^3, \,\, D = 2 , \label{so2D} \\
&& \delta\Omega (\bm B) = \frac{g^2 B^4}{3 \pi^2 v_F^3} \ln\left|\frac{2 B}{\Lambda}\right| \nonumber \\
&& \hspace{70pt} \times  \int\limits_{S_2} \frac{d \bm n}{4\pi} \left|\frac{\bet_{SO} (\bm n) \cdot \bm b}{\beta_{SO}(\bm n)}\right|^4  , \,\, D = 3 , \label{so3D} 
\end{eqnarray}
where $\bm b = \bm B / B$ is the unit vector along $\bm B$.
We neglected the term $\ln|\bet_{SO}(\bm n) \cdot \bm b / \beta_{SO}(\bm n)|$ in Eq.~(\ref{so3D})
because it just slightly renormalizes the regular $B^4$ term.
Here it is convenient to introduce the angular form-factor $F_D(\bm b)$ which depends 
on the direction $\bm b$ of the magnetic field 
and on the SO splitting:
\begin{eqnarray}
&& F_D (\bm b) = \int\limits_{S_{D - 1}} \frac{d \bm n}{S_{D - 1}} \left|\frac{\bet_{SO} (\bm n) \cdot \bm b}{\beta_{SO}(\bm n)}\right|^{D + 1} ,  \label{Fd}
\end{eqnarray}
where $S_{D - 1}$ is the area of a unit $(D-1)$-dimensional sphere.

The form-factors $F_D (\bm b)$
can only be positive or zero, see Eq.~(\ref{Fd}).
If we demand 
$F_D (\bm b) = 0$ for any unit vector $\bm b$,
it is equivalent to say that $\bet_{SO}(\bm n) \cdot \bm b = 0$ 
for any $\bm b$ and also $\beta_{SO}(\bm n) \ne 0$ from Eq.~(\ref{bet}).
As this is clearly impossible,
we conclude that $F_D(\bm b)$
can never vanish for all unit vectors $\bm b$
even at arbitrary SO splitting $\bet_{SO}(\bm n)$.
Therefore,
the non-analyticity with respect to $\bm B$
cannot be cut by any SO splitting neither in 2DEG nor in 3DEG.

Nevertheless, the SO splitting is important because it
leads to strong anisotropy
of the non-analytic term, see Eq.~(\ref{sogen}),
which is described by the form-factor $F_D(\bm b)$.
If we extrapolate this result to the vicinity of a FQPT, 
we conclude that
the direction of spontaneous magnetization
must coincide with the maximum of $F_D(\bm b)$.
In particular, 
we predict a first-order Ising FQPT
in electron gas with a general SO splitting which breaks the spin rotational symmetry
down to $\mathbb{Z}_2$.

As an example, we consider a 2DEG with Rashba and Dresselhaus SO splittings:
\begin{eqnarray}
&& \bet_{SO} (\phi)  \nonumber \\
&& = \left((\alpha_D + \alpha_R) k_F \sin \phi, (\alpha_D - \alpha_R) k_F \cos\phi, 0\right) ,
\end{eqnarray}
where the $x$ and $y$ axes correspond to the $[110]$ and $[1\overline{1}0]$ crystallographic directions,
$\alpha_R$ and $\alpha_D$ are the Rashba and the Dresselhaus coupling constants, respectively.
The qualitative picture of the SO-split Fermi surfaces
is shown in Fig.~\ref{fig:SOres}(a).
It is more convenient to introduce the following SO couplings:
\begin{eqnarray}
&& a_\pm \equiv \left(\alpha_R \pm \alpha_D\right) k_F .
\end{eqnarray}
Then we find the angular form-factor $F_2(\bm b)$, see Eqs.~(\ref{so2D}), (\ref{Fd}):
\begin{eqnarray}
&& F_2(\bm b)
= \int\limits_0^\pi \frac{d\phi}{\pi} \frac{\left|a_+ b_x \sin \phi \, - a_- b_y\cos \phi \, \right|^3}{\left(a_+^2 \sin^2 \phi + a_-^2 \cos^2 \phi\right)^{\frac{3}{2}}} , \label{Fphi}
\end{eqnarray}
where $\bm b = \bm B / B$ is the unit vector along $\bm B$.
We want to identify the directions $\bm b^*$ where 
$F_2(\bm b^*)$ is maximal.
It is clear that all such directions have $b^*_z = 0$.
Then $b_x^*$ and $b_y^*$ can be parametrized by a single angle $\Psi$:
\begin{eqnarray}
&& b_x^* = \cos \Psi, \,\, b_y^* = \sin \Psi .
\end{eqnarray}
The integral in Eq.~(\ref{Fphi})
is quite cumbersome but elementary:
\begin{widetext}
\begin{eqnarray}
&& \frac{2}{3} F_2(\zeta, \Psi) = - \frac{\zeta \cos\left(2 \Psi\right)}{\zeta^2 - 1}  + \frac{\zeta \cos \Psi}{(\zeta^2 - 1)^{\frac{3}{2}}} \left(\zeta^2 \cos^2 \Psi - 3 \sin^2 \Psi\right)
\arctan\left(\sqrt{\zeta^2 - 1} \cos \Psi\right)
\nonumber \\
&& + \frac{\sin \Psi}{(\zeta^2 - 1)^{\frac{3}{2}}} \left(\sin^2 \Psi - 3 \zeta^2 \cos^2 \Psi\right)
\ln\left(\frac{\sqrt{\zeta^2 \cos^2 \Psi + \sin^2 \Psi}}{\sqrt{\zeta^2 - 1} \sin \Psi + \zeta}\right) , \,\, \zeta \equiv \left|\frac{a_+}{a_-}\right| = \left|\frac{\alpha_R + \alpha_D}{\alpha_R - \alpha_D}\right| > 1 . \label{Fpsi}
\end{eqnarray}	
\end{widetext}
We added $\zeta$ as additional argument of $F_2(\bm b)$ for convenience.
Equation~(\ref{Fpsi}) is true only if $\zeta > 1$.
If $\zeta < 1$, we use the following identity:
\begin{eqnarray}
&& F_2(\zeta, \Psi) = F_2\left(\frac{1}{\zeta}, \frac{\pi}{2} - \Psi\right) . \label{Fsym}
\end{eqnarray}
The extremal values of the $\pi$-periodic function $F_2(\zeta, \Psi)$ 
correspond to $\Psi = 0$ and $\Psi = \pi/2$:
\begin{eqnarray}
&& F_2(\zeta, 0) = \frac{3}{2} \left[\frac{\zeta^3 \arctan\left(\sqrt{\zeta^2 - 1}\right)}{(\zeta^2 - 1)^{\frac{3}{2}}} -\frac{\zeta}{\zeta^2 - 1} \right] , \\
&& F_2\left(\zeta, \frac{\pi}{2}\right) = \frac{3}{2} \left[\frac{\zeta}{\zeta^2 - 1} -\frac{\ln\left(\zeta + \sqrt{\zeta^2 - 1}\right)}{(\zeta^2 - 1)^{\frac{3}{2}}} \right] , 
\end{eqnarray}
where $\zeta > 1$.
It is straightforward to see that at $\zeta > 1$
the maximum of $F_2(\zeta, \Psi)$ corresponds to $\Psi = 0$.
If $\zeta < 1$, we use Eq.~(\ref{Fsym})
and find that the maximum corresponds to $\Psi = \pi/2$.
If these calculations are extrapolated to 
the vicinity of the FQPT, 
we predict an Ising ferromagnetism 
in 2DEG with Rashba and Dresselhaus SO splitting.
The direction of spontaneous magnetization
here coincides
with the spin quantization axis 
of the states that are maximally split by the SO coupling,
namely, along $[110]$ ($[1\overline{1}0]$)
if $\alpha_R$ and $\alpha_D$ have the same (opposite) signs.
The case $\zeta = 1$ is realized when either Rashba or Dresselhaus SO splitting is zero,
in this case all in-plane directions are equivalent
which corresponds to the easy-plane ferromagnet, the result predicted in Refs.~\cite{zak1,zak2}.

\section{Strong interaction regime}

In this section we consider the electron gas 
with strong electron-electron interaction, i.e.
$g \gg 1$ or $k_F a_B \ll 1$, see Eq.~(\ref{coup2}).
In this section we concentrate 
on the case of SO splitting with 
small magnetic field $B \ll \beta$,
where $\beta$ is a characteristic 
SO splitting at the Fermi surface.
In this case 
we already know that only the scattering processes that 
are schematically shown in Fig.~\ref{fig:SOres}(b)
contribute to the non-analytic correction
in the regime of weak interaction.
In preceding sections we already discussed
that the scattering processes are nearly collinear in order 
to support the {\it local nesting},
see the paragraph after Eq.~(\ref{sogen}).
Here we consider the effective interaction Hamiltonian
whose matrix elements are only given by these 
processes, see Fig.~\ref{fig:SOres}(b).
Note that the standard backscattering 
within the same band is forbidden
due to Eq.~(\ref{mso}).
The momentum transfer in the diagram in Fig.~\ref{fig:SOres}(b) is about $2 k_F$,
so we can use the approximation of contact interaction, see Eq.~(\ref{contact}).
The only difference from the previous sections is that here
we consider the regime of strong electron-electron interactions $g \gg 1$,
i.e. we have to account for the processes in Fig.~\ref{fig:SOres}(b) 
fully self-consistently.

There is one more reason why we consider the 
case of finite SO splitting here.
As we found earlier, the SO splitting
results in strong anisotropy of the non-analytic terms, see Eq.~(\ref{sogen}).
For the case of general SO splitting
which breaks the spin rotational symmetry 
down to $\mathbb{Z}_2$,
the magnetic order parameter is Ising,
i.e. all fluctuation modes of the order parameter are gapped
and can be neglected even close to the FQPT.
At the same time, the resonant scattering processes
shown in Fig.~\ref{fig:SOres}
result in the negative non-analytic terms
in $\Omega (\bm B)$ and, thus,
destabilize the FQCP.
We show that the non-analyticity is enhanced parametrically in the limit of strong interaction $g \gg 1$.
This effect can be measured experimentally 
from the strongly non-analytic magnetic field dependence of the spin susceptibility close to the FQPT.

\subsection{Self-consistent Born approximation}

Next, we apply the strategy that 
we used before in Ref.~\cite{miserev}
where we predicted non-Fermi-liquid phases
(they correspond to certain magnetic quantum critical points)
in strongly interacting Fermi gases 
with multiple Fermi surfaces.
In fact, the case that we consider here
is very similar to a special case of Ref.~\cite{miserev}.
One difference here is that the Fermi surfaces are not spherical due to the anisotropic spin splitting.
Another difference here is that we analyze the stability 
of a FQCP via considering the non-analytic terms in the thermodynamic potential.

\begin{figure}[t]
	\centering
	\includegraphics[width=0.9\columnwidth]{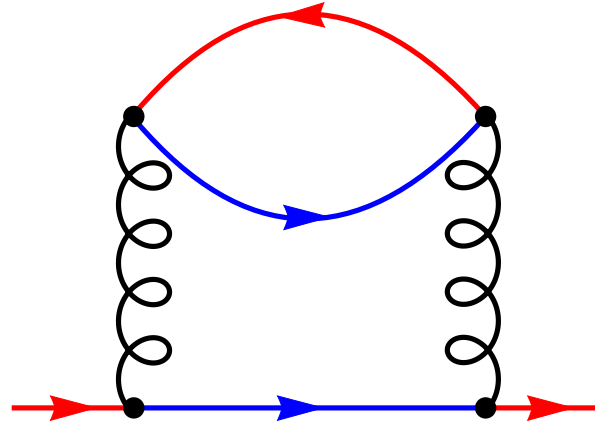}
	\caption{ 
		The lowest order diagram for the self-energy that is
		constructed from the effective Hamiltonian with the matrix elements 
		shown in Fig.~\ref{fig:SOres}(b).
		Here red (blue) color corresponds to 
		$\sigma = +1$ ($\sigma = -1$)
		Fermi surface.
		Inversion of the colors yields 
		the self-energy for $\sigma = -1$ electrons.
		The wiggly lines correspond to the 
		contact interaction defined in  Eq.~(\ref{contact}).
		We use this diagram for the self-consistent Born approximation.
	}
	\label{fig:sig}
\end{figure}

Here we consider the effective interaction
with non-zero matrix elements
that are given only by the 
diagram in Fig.~\ref{fig:SOres}(b) and its conjugate.
The lowest order diagram to the 
self-energy 
coming from such an effective interaction
is shown in Fig.~\ref{fig:sig}.
If we apply the perturbation theory and calculate the
corresponding non-analytic correction to $\Omega$,
we will restore Eq.~(\ref{sogen}).
In this section we want to go beyond  perturbation theory.
As a first step towards this goal, we treat 
the diagram in Fig.~\ref{fig:sig}
self-consistently, i.e. we dress all electron Green functions by corresponding self-energies:
\begin{eqnarray}
&& G_\sigma (i \omega, \bm p) \equiv G_\sigma (i \omega, \delta p, \bm n) \nonumber \\
&& \hspace{44pt} = \frac{|\sigma, \bm n\rangle \langle \sigma, \bm n|}{i \omega - v_F \, \delta p - \Sigma_\sigma (i \omega, \delta p , \bm n)} , \label{Gdressed}
\end{eqnarray}
where $\bm p = \bm k_\sigma + \bm n \, \delta p$, see Fig.~\ref{fig:fsexp}(a),
$\bm k_\sigma$ is the projection of $\bm p$ 
onto the Fermi surface $\mathcal{FS}_\sigma$,
$\bm n$ is the outward normal to $\mathcal{FS}_\sigma$ at $\bm k_\sigma$,
$\delta p$ is extended to the interval $(-\infty, \infty)$,
$\omega$ is the fermionic Matsubara frequency,
and $|\sigma, \bm n\rangle$
is given by Eq.~(\ref{psip}),
$v_F$ is the Fermi velocity at zero spin splitting.
In Eq.~(\ref{Gdressed})
we use that 
the interband backscattering does not alter the
single-particle spinors
because the backscattering matrix elements 
 equal  unity in absolute value, $|M_{+-}(\pm \bm n)| = 1$,
see Eq.~(\ref{mso}).
So far, we only dress the electron Green functions
leaving the effective interaction as contact interaction, see Eq.~(\ref{contact}),
and also neglecting the interaction vertex corrections.
This kind of approximations are usually referred to as
self-consistent Born approximations (SCBA).
The SCBA self-energy shown in Fig.~\ref{fig:sig} then reads:
\begin{eqnarray}
&& \Sigma_\sigma(z) = u^2 P_{-\sigma \sigma}(z) G_{-\sigma} (z) , \label{scba}
\end{eqnarray}
where $P_{\sigma\sigma'}(z)$ is the particle-hole bubble:
\begin{eqnarray}
&& P_{\sigma \sigma'}(z) = -{\rm Tr} \left\{G_\sigma(z) G_{\sigma'}(-z) \right\} . \label{Pdressed}
\end{eqnarray}
Note that the Green functions in 
Eqs.~(\ref{scba}) and (\ref{Pdressed})
are dressed by corresponding self-energies,
see Eq.~(\ref{Gdressed}).

Here we require the asymptotics of the dressed Green function
$G_\sigma (\tau, \bm r)$ at $\tau \gg 1/E_F$
and $r \gg \lambda_F$:
\begin{eqnarray}
&& G_\sigma (\tau, \bm r) \approx \left(\frac{1}{\lambda_F r}\right)^{\frac{D - 1}{2}} \left[e^{i \left(k_\sigma(\bm n_{\bm r}) r - \vartheta \right)} \mathcal{G}_\sigma (\tau, r, \bm n_{\bm r}) \right. \nonumber \\
&& \hspace{10pt} \left. + e^{-i \left( k_\sigma(-\bm n_{\bm r}) r - \vartheta \right)} \mathcal{G}_\sigma (\tau, -r, -\bm n_{\bm r})\right] , \,\,\, \bm n_{\bm r} = \frac{\bm r}{r} , \label{G} \\
&& \mathcal{G}_\sigma (\tau, x, \bm n) = T \sum\limits_{\omega} \int\limits_{-\infty}^\infty \frac{d \delta p}{2 \pi} e^{i \delta p x - i \omega \tau} G_\sigma(i \omega, \delta p, \bm n), \label{g} 
\end{eqnarray}
where $\vartheta$ is given by Eq.~(\ref{vartheta1}),
$\lambda_F$ is the Fermi wavelength, and
$\mathcal{G}_\sigma (\tau, x, \bm n)$
is the one-dimensional Fourier transform of 
$G_\sigma (i \omega, \delta p, \bm n)$, see Eq.~(\ref{Gdressed}).
The summation over Matsubara frequencies $\omega$
in Eq.~(\ref{g})
corresponds to finite temperatures $T > 0$.
If we neglect the self-energy in Eq.~(\ref{Gdressed}),
then Eq.~(\ref{G}) transforms into Eq.~(\ref{G0})
that we derived for the free electron Green function.
As the derivation of Eq.~(\ref{G})
is similar 
to the derivation of Eq.~(\ref{G0}) in many aspects,
we refer to Appendix for details.

First, we separate the spinors
using Eqs.~(\ref{Gdressed}) and (\ref{g}):
\begin{eqnarray}
&& \mathcal{G}_\sigma (\tau, x, \bm n) = |\sigma, \bm n\rangle \langle \sigma, \bm n| g_\sigma (\tau, x , \bm n) , \label{gtx}
\end{eqnarray}
where $g_\sigma (\tau, x, \bm n)$ is a scalar 1D Green function:
\begin{eqnarray}
&& g_\sigma (\tau, x, \bm n) \nonumber \\
&& \hspace{10pt} = T\sum\limits_{\omega} \int\limits_{-\infty}^\infty \frac{d\delta p}{2\pi} \frac{e^{i \delta p x - i \omega \tau}}{i \omega - v_F \delta p - \Sigma_\sigma (i \omega, \delta p, \bm n)} . \label{gdressed}
\end{eqnarray}
Using Eq.~(\ref{G}) in Eq.~(\ref{Pdressed}),
we find the asymptotics of the particle-hole bubble $P_{-\sigma \sigma}(z)$:
\begin{widetext}
\begin{eqnarray}
&& P_{-\sigma \sigma} (\tau, \bm r) \approx - \left(\frac{1}{\lambda_F r}\right)^{D - 1}  
\left[ e^{-2 i \vartheta} e^{i r (k_{-\sigma} (\bm n_{\bm r}) + k_{\sigma}(-\bm n_{\bm r}))} g_{-\sigma}(\tau, r, \bm n_{\bm r}) g_\sigma (-\tau, r, -\bm n_{\bm r})
\right. \nonumber \\
&& \hspace{110pt} \left.
 + e^{2 i \vartheta} e^{-i r (k_{-\sigma} (-\bm n_{\bm r}) + k_{\sigma}(\bm n_{\bm r}))} g_{-\sigma}(\tau, -r, -\bm n_{\bm r}) g_\sigma (-\tau, -r, \bm n_{\bm r})
\right] . \label{Pmss}
\end{eqnarray}
\end{widetext}
Here we used the matrix elements 
at $\bm B = 0$, see Eq.~(\ref{mso}),
because we calculate the leading non-analytic contribution.
Then we substitute Eq.~(\ref{Pmss}) into Eq.~(\ref{scba})
and use Eq.~(\ref{G}) for the Green function
to represent the self-energy in the following form:
\begin{eqnarray}
&& \hspace{-15pt} \Sigma_\sigma (\tau, \bm r) \approx \left(\frac{1}{\lambda_F r}\right)^{\frac{D - 1}{2}} \nonumber \\
&& \hspace{-15pt} \times \left[ e^{i (k_\sigma (\bm n_{\bm r}) r - \vartheta)} \mathcal{S}_\sigma (\tau, r, \bm n_{\bm r}) \, |\sigma, \bm n_{\bm r}\rangle \langle \sigma, \bm n_{\bm r}| \right. \nonumber \\
&& \hspace{-15pt} + \left. e^{-i (k_\sigma (-\bm n_{\bm r}) r - \vartheta)} \mathcal{S}_\sigma (\tau, -r, -\bm n_{\bm r}) \, 
|\sigma, -\bm n_{\bm r}\rangle \langle \sigma, -\bm n_{\bm r}|
\right] , \label{Sigas} 
\end{eqnarray}
where $\mathcal{S}_\sigma(\tau, x, \bm n)$ 
is the following function:	
\begin{eqnarray}
	&& \mathcal{S}_\sigma (\tau, x, \bm n) = - \frac{ u^2	e^{-i \sigma \Delta (\bm n) x}}{\left|\lambda_F x\right|^{D - 1}}
	\nonumber \\
	&& \hspace{20pt} \times 
	g_{-\sigma}(\tau, x, \bm n) g_{-\sigma} (\tau, -x, -\bm n) g_\sigma (-\tau, x, - \bm n) , \label{S} \\
	&& \Delta (\bm n) = k_+ (\bm n) - k_+(-\bm n) + k_-(-\bm n) - k_-(\bm n) \nonumber \\
	&& \hspace{20pt} \approx \frac{2}{v_F} \left(\beta (\bm n) - \beta (-\bm n)\right) \approx \frac{4}{v_F} \frac{\bet_{SO}(\bm n) \cdot \bm B}{\beta_{SO} (\bm n)} . \label{delta}
\end{eqnarray}
Here, we used Eqs.~(\ref{ksig}) and  (\ref{db})
to simplify $\Delta (\bm n)$.
Notice that Eq.~(\ref{Sigas}) has exactly the same form as Eq.~(\ref{G})
for the Green function.
This means that $\mathcal{S}_\sigma (\tau, x, \bm n)$ is just 
a 1D Fourier transform of the self-energy:
\begin{eqnarray}
&& \hspace{-10pt} \mathcal{S}_\sigma (\tau, x, \bm n) = T \sum\limits_{\omega} \int\limits_{-\infty}^\infty \frac{d\delta p}{2\pi} e^{i \delta p \, x - i \omega \tau} \Sigma_\sigma (i \omega, \delta p, \bm n) . \label{S1D}
\end{eqnarray}
Equation~(\ref{S}) provides the self-energy as a function of $g_\sigma (\tau, x, \bm n)$
which is itself connected to the self-energy
via Eq.~(\ref{gdressed}).

\subsection{Limit of strong interaction}

We expect a FQPT at large value of the dimensionless
interaction parameter $g \gg 1$.
In this regime,
the self-energy dominates
over the single-particle terms in the Green function,
so we can simplify Eq.~(\ref{gdressed}):
\begin{eqnarray}
&& g_\sigma (\tau, x, \bm n) \approx T\sum\limits_{\omega} \int\limits_{-\infty}^\infty \frac{d\delta p}{2\pi} \frac{e^{i \delta p x - i \omega \tau}}{- \Sigma_\sigma (i \omega, \delta p, \bm n)} . \label{gstrong}
\end{eqnarray}
Taking the convolution of $\mathcal{S}_\sigma (\tau, x, \bm n)$
and $g_\sigma (\tau, x, \bm n)$
given by Eqs.~(\ref{S1D}) and (\ref{gstrong}),
respectively,
we find:
\begin{eqnarray}
&& -\delta(\tau) \delta (x) \nonumber \\
&& \hspace{10pt} = \int d\tau' dx' \, g_\sigma (\tau - \tau', x - x', \bm n) \mathcal{S}_\sigma (\tau', x', \bm n) . \label{dyson}
\end{eqnarray}
Substituting Eq.~(\ref{S}) for the self-energy 
in Eq.~(\ref{dyson}),
we find the self-consistent equation for
the reduced Green function $g_\sigma (\tau, x, \bm n)$
in the limit of strong electron-electron interaction:
\begin{widetext}
	\begin{eqnarray}
	&& \delta(\tau) \delta(x) = u^2 \int d\tau' dx' \, g_\sigma (\tau - \tau', x - x', \bm n) g_\sigma(-\tau', x', -\bm n) g_{-\sigma} (\tau', x', \bm n) g_{-\sigma} (\tau', - x', -\bm n) \frac{e^{- i \sigma \Delta (\bm n) x'}}{|\lambda_F x'|^{D - 1}} . \label{geq}
	\end{eqnarray}
\end{widetext}

\subsection{Solutions at \mbox{\boldmath{$B = 0$}}}

First, we analyze the solutions of Eq.~(\ref{geq}) 
at zero magnetic field $\bm B = 0$
when $\Delta (\bm n) = 0$, see Eq.~(\ref{delta}).
First of all, we can drop the dependence of the reduced Green function on $\bm n$,
because there is no explicit dependence on $\bm n$ in Eq.~(\ref{geq})
if $\Delta(\bm n) = 0$:
\begin{eqnarray}
&& g_\sigma (\tau, x, \bm n) = g_\sigma (\tau, x) .
\end{eqnarray}
Second, we observe
that Eq.~(\ref{geq})
is free from any energy or momentum scales
which results in the scale invariance 
(conformal symmetry)
of Eq.~(\ref{geq})
with respect to independent 
reparametrization of time and coordinate:
\begin{eqnarray}
&& \tau \to \ell_1  \tau, \,\, x \to \ell_2 x ,
\end{eqnarray}
where $\ell_1$ and $\ell_2$ are arbitrary real numbers.
Applying this reparametrization to Eq.~(\ref{geq}), one can show
that
the Green function with rescaled 
coordinates $g_\sigma (\ell_1 \tau, \ell_2 x)$
is just proportional to $g_\sigma (\tau, x)$:
\begin{eqnarray}
&& \hspace{-10pt} g_\sigma (\ell_1 \tau, \ell_2 x) = \frac{g_\sigma (\tau, x)}{|\ell_1|^{2 d_1} |\ell_2|^{2 d_2}} , \,\, d_1 = \frac{1}{4}, \,\, d_2 = \frac{3 - D}{8} , \label{scale}
\end{eqnarray}
where $d_1$ and $d_2$ are the temporal and 
spatial scaling dimensions.
Equation~(\ref{scale}) implies 
the power-law scaling of the 
Green function $g_\sigma(\tau, x)$ with respect to time and coordinate:
\begin{eqnarray}
&& g_\sigma (\tau, x) \propto \frac{1}{|\tau|^{2 d_1}} \frac{1}{|x|^{2 d_2}} .
\end{eqnarray}
Thus, we see that
we can separate the variables
and represent Eq.~(\ref{geq})
at $\Delta(\bm n) = 0$
as the product of two independent Dyson equations:
\begin{eqnarray}
&& g_\sigma (\tau, x) = \mathfrak{g}(\tau) \gamma_\sigma (x) , \label{sep} \\
&& \delta (\tau) = \int\limits_{-\infty}^\infty d \tau' \, \mathfrak{g}(\tau - \tau') \mathfrak{g}(-\tau') \mathfrak{g}^2(\tau'), \label{gtau} \\
&& \delta(x) = u^2 \int\limits_{-\infty}^\infty dx' \, \gamma_\sigma (x - x') \frac{\gamma_\sigma (-x') \gamma^2_{-\sigma} (x')}{|\lambda_F x'|^{D - 1}} . \label{gammax}
\end{eqnarray}

Equation~(\ref{gtau}) is the well-known
Dyson equation of the Sachdev-Ye-Kitaev (SYK) model
\cite{ye,parcollet,sachdev}
which can be applied to describe 
the strange metal phase in cuprates \cite{sachdev,chowdhury}.
In Ref.~\cite{miserev}
we found that this model can emerge
in strongly interacting
electron systems
with multiple Fermi surfaces.
The zero temperature solution of this equation is the following:
\begin{eqnarray}
&& \mathfrak{g}(\tau) = \frac{1}{(4 \pi)^{\frac{1}{4}}} \frac{{\rm sgn}(\tau)}{|\tau|^{\frac{1}{2}}} . \label{syk}
\end{eqnarray}
In fact, this solution can be easily 
generalized to the case of finite temperature $T$:
\begin{eqnarray}
&& \mathfrak{g}(\tau) = \frac{{\rm sgn}(\tau)}{(4 \pi)^{\frac{1}{4}}} \left|\frac{\pi T}{\sin (\pi T \tau)}\right|^{\frac{1}{2}} , \label{sykT}
\end{eqnarray}
where $\tau \in (-1/T, 1/T)$.
The factor ${\rm sgn} (\tau)$ 
is needed here to satisfy 
the antiperiodicity of the electron Green function:
\begin{eqnarray}
&& G_\sigma \left(\tau + \frac{1}{T}, \bm r\right) = - G_\sigma (\tau, \bm r) .
\end{eqnarray}

Equation~(\ref{gammax}) is 
similar to the SYK equation~(\ref{gtau})
and can be solved in terms of the power-law functions:
\begin{eqnarray}
&& \gamma_\sigma (x) = C_D \frac{i \, {\rm sgn}(x)}{|x|^{2 d_2}} , \,\, d_2 = \frac{3 - D}{8} , \label{gamma}\\
&& C_D = \left[\frac{D + 1}{8 \pi u^2} \lambda_F^{D - 1} \tan \left(\frac{\pi}{2} \frac{3 - D}{4}\right) \right]^{\frac{1}{4}}. \label{C}
\end{eqnarray}
Here we assume that the proportionality coefficient $C_D$ 
is the same for both spin-split bands, so $\gamma_\sigma (x)$ is independent of $\sigma$ in this case.
This assumption is reasonable because the spin splitting is much smaller than the Fermi energy.

Substituting Eqs.~(\ref{sykT}) and (\ref{gamma})
into Eq.~(\ref{G}), we find the asymptotic form of the SCBA Green function:
\begin{eqnarray}
&& G_\sigma (\tau, \bm r) \approx \frac{C_D}{\lambda_F^{\frac{D - 1}{2}}}  \frac{\mathfrak{g}(\tau)}{r^{\frac{D + 1}{4}}} \left[e^{i \left(k_\sigma (\bm n_{\bm r}) r - \vartheta + \frac{\pi}{2}\right)} |\sigma, \bm n_{\bm r}\rangle \langle\sigma, \bm n_{\bm r}|\right. \nonumber \\
&& \hspace{20pt} +  \left. e^{-i \left(k_\sigma (-\bm n_{\bm r}) r - \vartheta + \frac{\pi}{2}\right)} |\sigma, - \bm n_{\bm r}\rangle \langle\sigma, - \bm n_{\bm r}| \right] , \label{Gcrit1}
\end{eqnarray}
where the $\pi/2$ phase is due to the $i \, {\rm sgn} (x)$ phase term in Eq.~(\ref{gamma}).
Equation~(\ref{Gcrit1}) has a certain phase freedom upon a small translation 
$r \to r + a$, $a \sim \lambda_F$.
Such a translation does not change the asymptotics of the power-law tail
because $r \gg \lambda_F$
but it results in the phase shift
in the oscillatory factors.
In other words, 
the phase cannot be determined self-consistently
from the long-range infrared limit.
Here we derive it from the particle-hole symmetry 
which is exact close to the Fermi surface,
so we
conclude that the phase has to be just $-\vartheta$.
The true asymptotics of the
particle-hole symmetric Matsubara Green function
with positively defined spectral function
then reads:
\begin{eqnarray}
&& G_\sigma (\tau, \bm r) \approx \frac{C_D}{\lambda_F^{\frac{D - 1}{2}}}  \frac{\mathfrak{g}(\tau)}{r^{\frac{D + 1}{4}}} \left[e^{i \left(k_\sigma (\bm n_{\bm r}) r - \vartheta \right)} |\sigma, \bm n_{\bm r}\rangle \langle\sigma, \bm n_{\bm r}|\right. \nonumber \\
&& \hspace{20pt} +  \left. e^{-i \left(k_\sigma (-\bm n_{\bm r}) r - \vartheta \right)} |\sigma, - \bm n_{\bm r}\rangle \langle\sigma, - \bm n_{\bm r}| \right]  . \label{Gcrit2}
\end{eqnarray}
The numerical coefficient $C_D > 0$
is given by Eq.~(\ref{C}).

Here we would like to comment on the case $D = 3$
because $C_D$ vanishes in this case.
In order to resolve this issue,
we use the dimensional regularization:
\begin{eqnarray}
&& C_D \approx \sqrt{\frac{\lambda_F}{4 u}}  \delta^{\frac{1}{4}}, \,\, D = 3 - \delta, \,\, \delta \to +0 .
\end{eqnarray}
Substituting it into Eq.~(\ref{gamma})
and forgetting about the phase factor (see the paragraph above),
we find:
\begin{eqnarray}
&& \gamma_\sigma (x) = \sqrt{\frac{\lambda_F}{4 u}} \left(\frac{|x|^\delta}{\delta}\right)^{-\frac{1}{4}} \approx \sqrt{\frac{\lambda_F}{4 u}} \left(\frac{1}{\delta} + \ln |x|\right)^{-\frac{1}{4}} .
\end{eqnarray}
Here $1/\delta$ comes from the ultraviolet 
scale $x \sim \lambda_F$,
so we can write $1/\delta = \ln p_0$, 
where $p_0 \sim k_F$
compensates the dimensionality of $x$:
\begin{eqnarray}
&& \gamma_\sigma (x) = \sqrt{\frac{\lambda_F}{4 u}} \frac{1}{\left(\ln|p_0 x|\right)^{\frac{1}{4}}} , \,\, p_0 \sim k_F .
\end{eqnarray}

As a summary, we provide the Matsubara Green function of 
strongly interacting 2D and 3D electron gases 
with arbitrary SO splitting:
\begin{widetext}
\begin{eqnarray}
&& G_\sigma (\tau, \bm r) \approx \left(\frac{3 (\sqrt{2} - 1) m^2 k_F}{\pi g^2}\right)^{\frac{1}{4}} \frac{e^{i \left(k_\sigma (\bm n_{\bm r}) r - \frac{\pi}{4} \right)} |\sigma, \bm n_{\bm r}\rangle \langle\sigma, \bm n_{\bm r}| + 
e^{-i \left(k_\sigma (-\bm n_{\bm r}) r - \frac{\pi}{4} \right)} |\sigma, -\bm n_{\bm r}\rangle \langle\sigma, -\bm n_{\bm r}|}{4 \pi r^{\frac{3}{4}}} \nonumber \\
&& \hspace{280pt} \times  \left|\frac{\pi T}{\sin (\pi T \tau)}\right|^{\frac{1}{2}} {\rm sgn} (\tau) , \,\, D = 2 , \label{grt2}\\
&& G_\sigma (\tau, \bm r) \approx \frac{k_F}{4 \pi^2} \left(\frac{4 \pi m^2}{g^2}\right)^{\frac{1}{4}} \frac{e^{i \left(k_\sigma (\bm n_{\bm r}) r - \frac{\pi}{2} \right)} |\sigma, \bm n_{\bm r}\rangle \langle\sigma, \bm n_{\bm r}| + 
	e^{-i \left(k_\sigma (-\bm n_{\bm r}) r - \frac{\pi}{2} \right)} |\sigma, -\bm n_{\bm r}\rangle \langle\sigma, -\bm n_{\bm r}|}{r \left|\ln(p_0 r)\right|^{\frac{1}{4}}} \nonumber \\
&& \hspace{280pt} \times \left|\frac{\pi T}{\sin (\pi T \tau)}\right|^{\frac{1}{2}} {\rm sgn} (\tau) , \,\, D = 3 , \label{grt3}
\end{eqnarray}
\end{widetext}
where $g$ is the dimensionless interaction coupling constant, see Eq.~(\ref{coup2}).
We also provide the Fourier transforms:
\begin{eqnarray}
&& G_\sigma (i \omega, \delta p, \bm n) = \left(\frac{3 \pi}{2}\right)^{\frac{1}{4}} \frac{(\sqrt{2} -1)^{\frac{3}{4}} \Gamma\left(\frac{3}{4}\right)}{2 \pi \sqrt {2 g E_F T}} \nonumber \\
&& \hspace{60pt} \times \left|\frac{k_F}{\delta p}\right|^{\frac{3}{4}} \frac{\Gamma \left(\frac{\omega}{2 \pi T} + \frac{1}{4}\right)}{\Gamma \left(\frac{\omega}{2 \pi T} + \frac{3}{4}\right)} , \,\, D = 2 , \label{gkw2} \\
&& G_\sigma (i \omega, \delta p, \bm n) =  \frac{\pi^{\frac{1}{4}}}{16 \sqrt {g E_F T}} \nonumber \\
&& \hspace{10pt} \times \frac{k_F}{|\delta p|}
\left(\ln\left|\frac{p_0}{\delta p}\right|\right)^{-\frac{5}{4}} \frac{\Gamma \left(\frac{\omega}{2 \pi T} + \frac{1}{4}\right)}{\Gamma \left(\frac{\omega}{2 \pi T} + \frac{3}{4}\right)} , \,\, D = 3 , \label{gkw3}
\end{eqnarray}
where $\omega = \pi T (2 n + 1)$ is the Matsubara frequency, with $n$ being an integer,
and, again, $p_0 \sim k_F$.

These calculations 
have been performed within the SCBA.
However, the emergent conformal symmetry 
makes this approximation exact,
see Ref.~\cite{miserev} for details.
In particular,
the interaction vertex correction and the dressing of the 
effective interaction 
only lead to the renormalization of the interaction parameter $g$,
while the scaling dimensions $d_1$ and $d_2$
remain the same.
In this paper we consider $g$ as a phenomenological parameter,
so its renormalizations are not important for us.
The only condition that is implied here is that 
$g \gg 1$ close to the FQPT, so we can apply the limit of strong interaction.

\subsection{Solutions at finite $\bm B$}

If $\bm B$ is finite, then Eq.~(\ref{geq})
contains an oscillatory term
making the self-energy irrelevant at 
large distance $x \gg 1/\Delta(\bm n)$,
so we expect Fermi liquid 
behavior
 at such distances.
At small distances $x \ll 1/\Delta(\bm n)$
the oscillatory term is not important
and the solutions, given by Eqs.~(\ref{grt2})--(\ref{grt3}),
are still valid.
Following this reasoning,
the Fourier transforms 
Eqs.~(\ref{gkw2})--(\ref{gkw3})
are valid if $\delta p \gg \Delta (\bm n)$,
where $\bm n$ here labels the direction of $\bm p$.

Another important limitation of Eqs.~(\ref{grt2})--(\ref{gkw3})
comes from the assumption that the self-energy
$\Sigma_\sigma (i \omega, \delta p, \bm n)$
dominates over
the single-particle terms $i \omega - v_F \delta p$
which we neglected in Eq.~(\ref{gstrong}).
All in all,
the non-Fermi liquid regime 
corresponds to the following constrained region:
\begin{eqnarray}
&&\hspace{-20pt} \Delta(\bm n) \ll  \delta p \ll k_F, \,\, w(\delta p) \ll \omega \ll W (\delta p) , \label{region} \\
&& \hspace{-20pt} w(\delta p) = \frac{E_F}{g} \left|\frac{\delta p}{k_F}\right|^{\frac{1}{2}} , \, W(\delta p) = g E_F \left|\frac{\delta p}{k_F}\right|^{\frac{3}{2}} , \, D = 2 , \label{const2} \\
&& \hspace{-20pt} w(\delta p) = \frac{E_F}{g} \left(\ln \left|\frac{k_F}{\delta p}\right|\right)^{-\frac{5}{2}} , \nonumber \\
&& \hspace{12pt} W(\delta p) = g E_F \left(\frac{\delta p}{k_F}\right)^2 \left(\ln \left|\frac{k_F}{\delta p}\right|\right)^{\frac{5}{2}}, \, D = 3 ,\label{const3}
\end{eqnarray}
where we used $p_0 \approx k_F$ for $D = 3$.
Here, it is required that $W(\delta p) \gg w(\delta p)$ 
which further constrains $\delta p$:
\begin{eqnarray}
&& \delta p \gg \delta p^* , \label{star} \\
&& \delta p^* = \frac{k_F}{g^2} , \, D = 2 , \\
&& \delta p^* = \frac{k_F}{g \left(\ln g\right)^{\frac{5}{2}}} , \, D = 3 .
\end{eqnarray}
As $g \gg 1$ close to the FQPT,
the scale $\delta p^*$ is very small.
Here we see that at the very deep infrared limit
$\delta p \lesssim \delta p^*$
the Fermi liquid is restored.
This also means that at $\Delta (\bm n) \ll \delta p^*$
the non-analyticity can be correctly estimated by the perturbative approach that we considered 
in the previous section.
The most interesting regime corresponds to 
$\Delta (\bm n) \gg \delta p^*$ 
where the dominant contribution to the thermodynamic potential comes from the
quantum states that are strongly affected by the interaction.
In the next section we calculate the thermodynamic potential 
$\Omega$
in this regime:
\begin{eqnarray}
&& \Delta (\bm n) \gg \delta p^* . \label{delstar}
\end{eqnarray}

\section{Thermodynamic potential}

In the preceding section 
we calculated the electron Green function
of 2DEG and 3DEG with arbitrary SO splitting in
the strongly interacting regime $g \gg 1$,
see Eqs.~(\ref{grt2})--(\ref{gkw3}),
and we also 
defined the crossover region with
the Fermi liquid,
see Eqs.~(\ref{region}) and (\ref{delstar}).
The Fermi liquid contribution to the 
non-analyticity is calculated in Sec.~IV.
Here we address the contribution
coming from the strongly correlated region
of the phase space, see Eqs.~(\ref{region}) and (\ref{delstar}).

In order to calculate the thermodynamic potential $\Omega$,
we consider the following energy functional $\Omega(\Sigma, G)$:
\begin{eqnarray}
&& \hspace{-10pt} \Omega (\Sigma, G) = - {\rm Tr} \ln \left(G_0^{-1} - \Sigma\right) - {\rm Tr} \left(\Sigma G\right) - \Phi (G) , \label{func}
\end{eqnarray}
where ${\rm Tr}$ is the trace over all indexes and
$\Phi(G)$ the Luttinger-Ward functional.
The saddle point of $\Omega(\Sigma, G)$ yields
the Dyson equation and the equation for the electron self-energy:
\begin{eqnarray}
&& 0 = \frac{\delta \Omega (\Sigma, G)}{\delta \Sigma_\sigma} = \left(G_0^{-1} -\Sigma_\sigma \right)^{-1} - G_\sigma , \\
&& 0 =  \frac{\delta \Omega (\Sigma, G)}{\delta G_\sigma} = -\Sigma_\sigma -  \frac{\delta \Phi (G)}{\delta G_\sigma} . \label{selfen}
\end{eqnarray}
Comparing Eq.~(\ref{scba}) with Eq.~(\ref{selfen}), we reconstruct the Luttinger-Ward functional:
\begin{eqnarray}
&& \Phi (G) = \frac{u^2}{2} \int dz \, \left[{\rm Tr} \left\{G_+(z) G_-(-z)\right\}\right]^2 . \label{LW}
\end{eqnarray}
The thermodynamic potential $\Omega$ is given by the saddle-point value of the 
functional $\Omega (\Sigma, G)$:
\begin{eqnarray}
&& \Omega = {\rm Tr} \ln G + 3 \Phi(G) , \label{om}
\end{eqnarray}
where $G$ is the Matsubara Green function.
Here we used that $\Sigma$ satisfies Eq.~(\ref{scba}), so that $-{\rm Tr}(\Sigma G) = 4 \Phi (G)$.

\subsection{The logarithmic term}

Let us first calculate ${\rm Tr} \ln G$:
\begin{eqnarray}
&& \hspace{-18pt} \Omega_1 \equiv {\rm Tr} \ln G = \sum\limits_\sigma T \sum\limits_{\omega} \int \frac{d \bm p}{(2 \pi)^D} \ln \left(G_\sigma (i \omega, \delta p)\right) . \label{om1}
\end{eqnarray}
Here we assume that Eq.~(\ref{delstar}) is satisfied,
so the quantum critical region 
is given by the constraints Eqs.~(\ref{region})--(\ref{const3}).

We start from the low temperature regime:
\begin{eqnarray}
&& T \ll W(\Delta(\bm n)) . \label{lowT}
\end{eqnarray}
In this case there are many Matsubara frequencies 
in the window $\omega \in (w(\Delta(\bm n)), W(\Delta(\bm n))$,
so we can approximate the sum over frequencies by an integral.
For the same reason, we can also expand the gamma functions in Eqs.~(\ref{gkw2})--(\ref{gkw3})
with respect to the large ratio $\omega/(2 \pi T)$
such that $\ln G_\sigma (i \omega, \delta p)$
takes the following form:
\begin{eqnarray}
&& \hspace{-20pt} \ln G_\sigma (i \omega, \delta p) = \frac{3}{4} \ln \left|\frac{k_F}{\delta p}\right| - \frac{1}{2} \ln |\omega| + {\rm const} , \, D = 2 , \label{lng2}\\
&& \hspace{-20pt} \ln G_\sigma (i \omega, \delta p) =  \ln \left|\frac{k_F}{\delta p}\right| - \frac{5}{4} \ln \ln \left|\frac{p_0}{\delta p}\right| \nonumber \\
&& \hspace{90pt} - \frac{1}{2} \ln |\omega| + {\rm const} , \, D = 3 . \label{lng3}
\end{eqnarray}
The sum over $\sigma$ in Eq.~(\ref{om1}) yields just a factor of $2$.
As $\ln G_\sigma(i \omega, \delta p)$ is an even function of $\omega$ and 
$\delta p$, we 
can integrate over the positive intervals,
i.e. $\delta p \in (\Delta(\bm n), k_F)$
and $\omega \in (0, W(\delta p))$.
Note that we extend the integration over $\omega$ 
all the way to zero frequency
because $W(\delta p) \gg T$ and $W(\delta p) \gg w(\delta p)$.
This allows us to represent Eq.~(\ref{om1}) in the following form:
\begin{eqnarray}
&& \Omega_1(\bm B) = 2 \int\limits_0^{2\pi} \frac{d\phi}{2\pi} 2 \int\limits_{\Delta(\phi)}^{k_F} \frac{k_F \, d\delta p}{2 \pi} 2 \int\limits_0^{W(\delta p)} \frac{d\omega}{2\pi} \nonumber \\
&& \hspace{60pt}\times \left(\frac{3}{4} \ln \left|\frac{k_F}{\delta p}\right| - \frac{1}{2} \ln |\omega|\right), \, D = 2, \label{om12} \\
&& \Omega_1(\bm B) = 2 \int \frac{d\bm n}{4\pi} 2 \int\limits_{\Delta(\bm n)}^{k_F} \frac{k_F^2 \, d\delta p}{2 \pi^2} 2 \int\limits_0^{W(\delta p)} \frac{d\omega}{2\pi} \nonumber \\
&& \hspace{10pt}\times \left(\ln \left|\frac{k_F}{\delta p}\right| - \frac{5}{4} \ln \ln \left|\frac{p_0}{\delta p}\right| - \frac{1}{2} \ln |\omega|\right), \, D = 3 . \label{om13}
\end{eqnarray}
Next, we take the integral over $\omega$
with logarithmic accuracy.
In particular, we neglect the $\ln\ln|\delta p|$ term in the 3D case compared to the $\ln|\delta p|$ term:
\begin{eqnarray}
&& \Omega_1(\bm B) = \frac{2 k_F^2}{\pi^2} \int\limits_0^{2\pi} \frac{d\phi}{2\pi} \int\limits_{\Delta(\phi)}^{k_F} \frac{d\delta p}{k_F} W(\delta p)\nonumber \\
&& \hspace{40pt}\times \left(\frac{3}{4} \ln \left|\frac{k_F}{\delta p}\right| - \frac{1}{2} \ln |W(\delta p)|\right), \, D = 2, \label{om12k} \\
&& \Omega_1(\bm B) = \frac{2 k_F^3}{\pi^3} \int \frac{d\bm n}{4\pi}  \int\limits_{\Delta(\bm n)}^{k_F} \frac{d\delta p}{k_F} W(\delta p) \nonumber \\
&& \hspace{40pt} \times\left(\ln \left|\frac{k_F}{\delta p}\right| - \frac{1}{2} \ln |W(\delta p)|\right), \, D = 3 . \label{om13k}
\end{eqnarray}
Then substituting $W(\delta p)$, see Eqs.~(\ref{const2})--(\ref{const3}),
we find:
\begin{eqnarray}
&&  \Omega_1 (\bm B) = \frac{3 k_F^2 g E_F}{\pi^2} \int\limits_0^{2\pi}\frac{d\phi}{2\pi} \nonumber 
 \\
&& \hspace{50pt} \times
 \int\limits_{\Delta(\phi)}^{k_F} \frac{d\delta p}{k_F} \left|\frac{\delta p}{k_F}\right|^{\frac{3}{2}} \ln \left|\frac{k_F}{\delta p}\right|, \, D = 2, \\
&&  \Omega_1 (\bm B) = \frac{4 k_F^3 g E_F}{\pi^3} \int\frac{d\bm n}{4\pi} 
\nonumber \\
&& \hspace{50pt}  \times
\int\limits_{\Delta(\bm n)}^{k_F} \frac{d\delta p}{k_F} \left|\frac{\delta p}{k_F}\right|^{2} \ln \left|\frac{k_F}{\delta p}\right|^{\frac{7}{2}}, \, D = 3 .
\end{eqnarray}
Calculating the integrals over $\delta p$ 
with the logarithmic accuracy 
and subtracting the part which is independent of $\Delta (\bm n)$,
we find the non-analytic contribution to the thermodynamic potential:
\begin{eqnarray}
&& \Omega_1 (\bm B) = -\frac{6 k_F^2 g E_F}{5 \pi^2} \nonumber \\
&& \hspace{40pt} \times \int\limits_0^{2\pi} \frac{d\phi}{2 \pi} \left|\frac{\Delta(\phi)}{k_F}\right|^{\frac{5}{2}} \ln\left|\frac{k_F}{\Delta(\phi)}\right|, \, D = 2, \label{lead2} \\
&& \Omega_1 (\bm B) = -\frac{4 k_F^3 g E_F}{3 \pi^3} \nonumber \\
&& \hspace{20pt} \times \int\frac{d\bm n}{4\pi} \left|\frac{\Delta(\bm n)}{k_F}\right|^3 \left(\ln\left|\frac{k_F}{\Delta(\bm n)}\right|\right)^{\frac{7}{2}},  \, D = 3 . \label{lead3}
\end{eqnarray}
Finally, we have to calculate the Luttinger-Ward
contribution to $\Omega$.

\subsection{Luttinger-Ward contribution}

Here we calculate the second part of the thermodynamic potential, see Eq.~(\ref{om}):
\begin{eqnarray}
&& \hspace{-20pt} \Omega_2(\bm B) = 3 \Phi(G) = \frac{3 u^2}{2} \!\!\int dz \left[{\rm Tr} \left\{G_+(z) G_-(-z)\right\}\right]^2 .
\end{eqnarray}
In this case it is more convenient to use 
the real space representation of the 
Green function, 
see Eqs.~(\ref{grt2})--(\ref{grt3}).
Here we will show that $\Omega_2(\bm B)$ 
yields a subleading correction compared to $\Omega_1(\bm B)$.

Using Eqs.~(\ref{grt2})--(\ref{grt3}),
we find:
\begin{eqnarray}
&& \left[{\rm Tr} \left\{G_+(z) G_- (-z)\right\}\right]^2 \approx \frac{3 (\sqrt{2} - 1) m^2 k_F}{2^7 \pi^5 g^2} \nonumber \\
&& \hspace{30pt} \times \frac{e^{i r \Delta(\bm n_{\bm r})}}{r^3} \left|\frac{\pi T}{\sin(\pi T \tau)}\right|^2 , \,  D = 2, \label{tr2} \\
&& \left[{\rm Tr} \left\{G_+(z) G_- (-z)\right\}\right]^2 \approx \frac{m^2 k_F^4}{2^5 \pi^7 g^2} \nonumber \\
&& \hspace{30pt} \times \frac{e^{i r \Delta(\bm n_{\bm r})}}{r^4 \ln(p_0 r)} \left|\frac{\pi T}{\sin(\pi T \tau)}\right|^2 , \,  D = 3, \label{tr3}
\end{eqnarray}
First, we have to translate the 
conditions given by Eqs.~(\ref{region})--(\ref{const3})
to constraints on $\tau$ and $r$.
We do it via the correspondence $\omega \sim 1/\tau$, $\delta p \sim 1/r$.
Thus, in our case the integration region is 
defined by the following constraints:
\begin{eqnarray}
&& \tau \gg \overline{\tau}(r) = \frac{1}{W(r^{-1})}, \,\, r \ll \frac{1}{\Delta (\bm n)} .
\end{eqnarray}
Here we show only the most important conditions.
As in the previous case, we consider small 
temperatures $T \ll W(\Delta(\bm n))$,
so we can actually use the $T = 0$ approximation.
First, we integrate over $\tau$:
\begin{eqnarray}
&& \int\limits_{\overline{\tau}(r)}^{\frac{1}{T}-\overline{\tau}(r)} d\tau \, \left|\frac{\pi T}{\sin(\pi T \tau)}\right|^2 \approx 2 \int\limits_{\overline{\tau}(r)}^\infty \frac{d\tau}{\tau^2} = \frac{2}{\overline{\tau}(r)} .
\end{eqnarray}
Substituting this in $\Omega_2(\bm B)$,
we find:
\begin{eqnarray}
&& \Omega_2(\bm B) = \frac{9 (\sqrt{2}- 1)}{16 \pi^2} g E_F k_F^2 \int\limits_0^{2\pi} \frac{d\phi}{2\pi} \nonumber \\
&& \hspace{70pt} \times \int\limits_0^\infty \frac{k_F dr}{(k_F r)^{\frac{7}{2}}} e^{i r \Delta(\phi)} , \, D = 2 , \label{omr2}\\
&& \Omega_2(\bm B) = \frac{3}{2 \pi^2} g E_F k_F^3 \int \frac{d\bm n}{4 \pi} \nonumber \\
&& \hspace{30pt} \times \int\limits_0^\infty \frac{k_F dr}{(k_F r)^4} \left(\ln(k_F r)\right)^{\frac{3}{2}} e^{i r \Delta(\bm n)} , \, D = 3 . \label{omr3}
\end{eqnarray}
We substituted $p_0 \approx k_F$ for $D = 3$ case.
Note that here we do not even have to use the condition
$r \ll 1/\Delta(\bm n)$
because it is effectively satisfied due to the 
oscillatory term $e^{i r \Delta(\bm n)}$.
The integral over $k_F r$ in Eq.~(\ref{omr2}) is calculated via Eq.~(\ref{int}).
As $\Delta(\phi + \pi) = -\Delta(\phi)$, see Eq.~(\ref{delta}),
then only the real part does not vanish after the integration over $\phi$.
The integral in Eq.~(\ref{omr3})
can be calculated similarly.
Here we calculate it with logarithmic accuracy:
\begin{eqnarray}
&& \int\limits_0^\infty \frac{k_F dr}{(k_F r)^{4+\delta}} \left(\ln(k_F r)\right)^{\frac{3}{2}} e^{i r \Delta(\bm n)} \nonumber \\
&& \approx \left(\ln\left|\frac{k_F}{\Delta(\bm n)}\right|\right)^{\frac{3}{2}} \left|\frac{\Delta(\bm n)}{k_F}\right|^{3 + \delta} \frac{i \, {\rm sgn}(\Delta (\bm n)) + \frac{\pi}{2} \delta}{6 \delta} ,
\end{eqnarray}
where we introduced some small $\delta \to 0$.
As $\Delta(-\bm n) = - \Delta(\bm n)$,
 the divergent $1/\delta$ part cancels out 
after the integration over $\bm n$,
so the physical contribution is finite.
In conclusion, we find $\Omega_2(\bm B)$:
\begin{eqnarray}
&& \Omega_2(\bm B) = \frac{3 (\sqrt{2}- 1)}{5 (2\pi)^{\frac{3}{2}}} g E_F k_F^2
\nonumber \\
&& \hspace{70pt} \times  \int\limits_0^{2\pi} \frac{d\phi}{2\pi} \left|\frac{\Delta(\phi)}{k_F}\right|^{\frac{5}{2}} , \, D = 2 , \label{om22}\\
&& \Omega_2(\bm B) = \frac{g E_F k_F^3}{8 \pi}   \nonumber \\
&& \hspace{20pt} \times \int \frac{d\bm n}{4 \pi}  \left|\frac{\Delta(\bm n)}{k_F}\right|^{3} \left(\ln\left|\frac{k_F}{\Delta(\bm n)}\right|\right)^{\frac{3}{2}} , \, D = 3 . \label{om23}
\end{eqnarray}
Indeed, compared to $\Omega_1(\bm B)$, see Eqs.~(\ref{lead2}) and (\ref{lead3}),
$\Omega_2(\bm B)$ is only subleading,
so we can neglect it.

Summing over, the leading non-analyticity 
in $\Omega$ in the strongly interacting regime 
$g \gg 1$ close to FQPT
is given by the $\Omega_1$ contribution, see Eqs.~(\ref{lead2}) and (\ref{lead3}).
Here we represent the answer using the 
angular form-factors
$\overline{F}_D(\bm b)$ 
that are different from the form-factors
$F_D(\bm b)$ that appear in the weak coupling regime, see Eq.~(\ref{Fd}):
\begin{eqnarray}
&& \hspace{-10pt}  \delta \Omega (\bm B) = \frac{6 k_F^2 g E_F}{5 \pi^2}
\left|\frac{2 B}{E_F}\right|^{\frac{5}{2}}
 \ln\left|\frac{2 B}{E_F}\right| \overline{F}_2(\bm b) , \, D = 2, \label{qpt2D} \\
&& \hspace{-10pt} \delta \Omega (\bm B) = -\frac{32 k_F^3 g |B|^3}{3 \pi^3 E_F^2}
\left|\ln\left|\frac{E_F}{2 B}\right|\right|^{\frac{7}{2}}
\overline{F}_3(\bm b) ,  \, D = 3 , \label{qpt3D} \\
&& \hspace{-10pt} \overline{F}_D (\bm b) = \int\limits_{S_{D - 1}} \frac{d \bm n}{S_{D - 1}} \left|\frac{\bet_{SO} (\bm n) \cdot \bm b}{\beta_{SO} (\bm n)}\right|^{\frac{D + 3}{2}} , \label{Fbar}
\end{eqnarray}
where $\bm b = \bm B/ B$.
First, we notice that the non-analytic terms are negative
and that they are parametrically larger
than the Fermi liquid non-analyticity,
see Eqs.~(\ref{so2D}) and (\ref{so3D}).
The angular form-factor $\overline{F}_D(\bm b)$ due to the SO splitting
is different from the one that we considered 
in the weak coupling regime, see Eq.~(\ref{Fd}),
but it still results in the same physical effect:
if the SO splitting breaks the spin rotational symmetry down
to $\mathbb{Z}_2$,
then the ferromagnetic ground state is Ising
and the magnetization axis is defined by the 
direction where the SO splitting
and, thus, $\overline{F}_D(\bm b)$ are maximal.

Equations~(\ref{qpt2D}) and (\ref{qpt3D})
are valid at very low temperatures
when $T \ll W(\Delta(\bm n))$.
This condition can be rewritten as follows:
\begin{eqnarray}
&& B \gg B^*(T) , \\
&& B^*(T) = E_F \left|\frac{T}{g E_F}\right|^{\frac{2}{3}} , \,\, D = 2,  \label{b2d} \\
&& B^* (T) = \frac{E_F}{\left(\ln|g E_F/T| \right)^{\frac{5}{4}}} \sqrt{\frac{T}{g E_F}} , \,\, D = 3 . \label{b3d}
\end{eqnarray}
If  $B \ll B^* (T)$,
then the electron self-energy 
at $\omega \sim B$ 
and $v_F \delta p \sim B$
is irrelevant,
so the regime $B \ll B^* (T)$
corresponds to the Fermi liquid regime
and the non-analyticity in $\Omega$
is described by Eqs.~(\ref{so2D})--(\ref{so3D}).
The regime $B \gg B^*(T)$ is quantum critical
and the non-analyticities are described by Eqs.~(\ref{qpt2D})--(\ref{qpt3D}).
The regime $B \sim B^*(T)$
corresponds to the crossover
between the Fermi-liquid
and the non-Fermi-liquid regimes.

In this section we treated the resonant backscattering processes
shown in Fig.~\ref{fig:SOres}(b)
non-perturbatively
and found that 
the non-analytic terms in $\Omega$
remain negative close to the FQPT and 
that they
are strongly enhanced in magnitude compared to the regime of weak interaction.
This means 
that the FQCP separating paramagnetic and ferromagnetic metals
is intrinsically unstable even in the presence of arbitrarily complex SO splitting.

\section{Spin susceptibility far away and close to FQPT}

The non-analytic corrections destabilizing
the FQCP separating ferromagnetic and paramagnetic phases
can be measured experimentally via the 
magnetic field dependence of the spin susceptibility $\chi_{ij} (\bm B)$
in the paramagnetic phase:
\begin{eqnarray}
&& \chi_{ij}(\bm B) = - \frac{\partial^2 \Omega (\bm B)}{\partial B_i \partial B_j} . \label{chi}
\end{eqnarray}
Deep inside the paramagnetic phase
where $g \ll 1$ or $\lambda_F \ll a_B$, see Eq.~(\ref{coup2}),
the non-analytic correction 
is given by Eqs.~(\ref{so2D})--(\ref{so3D}),
so the corresponding non-analytic corrections to the
spin susceptibility are the following:
\begin{eqnarray}
&&\hspace{-10pt} \delta \chi_{ij} (\bm B) = \frac{2 g^2 |B|}{\pi v_F^2} \kappa^{(2)}_{ij} (\bm b) , \,\, D = 2 , \label{chi2dweak}\\
&& \hspace{-10pt} \delta \chi_{ij} (\bm B) = - \frac{4 g^2 B^2}{\pi^2 v_F^3} \ln\left|\frac{2 B}{\Lambda}\right| \kappa^{(3)}_{ij} (\bm b) , \,\, D = 3,  \label{chi3dweak}\\
&& \hspace{-10pt} \kappa^{(D)}_{ij} (\bm b)\nonumber \\
&& \hspace{-5pt} = \int\limits_{S_{D - 1}} \frac{d \bm n}{S_{D - 1}} \left|\frac{\bet_{SO}(\bm n) \cdot \bm b}{\beta_{SO}(\bm n)}\right|^{D - 1} \frac{\beta_{SO}^i (\bm n) \beta_{SO}^j (\bm n)}{\beta^2_{SO}(\bm n)} , \label{kappa}
\end{eqnarray}
where $\Lambda \sim E_F$,
$\beta_{SO}^i (\bm n)$
is the $i^{{\rm th}}$ component of the vector $\bet_{SO} (\bm n)$.
These results are valid for $B \gg T$.
In the opposite regime $B \ll T$,
the non-analyticity is regularized by the temperature.
An important feature here
is the non-trivial angular dependence of the 
spin susceptibility 
on $\bm b$ due to the SO splitting, see the angular tensor $\kappa^{(D)}_{ij}(\bm b)$.
Measuring $\kappa^{(D)}_{ij}(\bm b)$
can resolve the 
angular dependence of the SO splitting.
However, it is not an easy task to measure 
the non-analytic corrections to the spin susceptibility
in the weak coupling regime 
because they are proportional to $g^2$,
where $g \ll 1$.

It is much more promising
to measure the non-analyticities in the spin susceptibility
when the paramagnetic phase is close to the FQPT,
such that $g \gg 1$ and we can use the results of Sec. VI,
see Eqs.~(\ref{qpt2D})--(\ref{qpt3D}):
\begin{eqnarray}
&& \delta \chi_{ij}(\bm B) \nonumber \\
&& \approx \frac{36 \sqrt{2}}{\pi^2} m g \ln\left|\frac{E_F}{2 B}\right| \sqrt{\frac{|B|}{E_F}} \, \overline{\kappa}_{ij}^{(2)} (\bm b) , \,\, D = 2 , \label{chistrong2d}\\
&& \delta \chi_{ij}(\bm B) \nonumber \\
&& \approx \frac{256 m g}{\pi^3 v_F} \left|\ln\left(\frac{E_F}{2 B}\right)\right|^{\frac{7}{2}} |B| \, \overline{\kappa}_{ij}^{(3)} (\bm b), \,\, D = 3 , \label{chistrong3d}\\
&& \overline{\kappa}^{(D)}_{ij} (\bm b)\nonumber \\
&& \hspace{-5pt} = \int\limits_{S_{D - 1}} \frac{d \bm n}{S_{D - 1}} \left|\frac{\bet_{SO}(\bm n) \cdot \bm b}{\beta_{SO}(\bm n)}\right|^{\frac{D - 1}{2}} \frac{\beta_{SO}^i (\bm n) \beta_{SO}^j (\bm n)}{\beta^2_{SO}(\bm n)} . \label{overkappa}
\end{eqnarray}
These equations are valid if $B \gg B^*(T)$,
where $B^*(T)$ is given by Eqs.~(\ref{b2d}) and (\ref{b3d}).
If $B \ll B^*(T)$,
then the electron self-energy is irrelevant 
on the scale $\omega \sim B$
and the non-analyticity is given by 
Eqs.~(\ref{chi2dweak}) and (\ref{chi3dweak}).
Here we see that at $B \gg B^*(T)$
the non-analytic corrections to the susceptibility
are strongly enhanced.
In particular, $ \delta \chi_{ij} (\bm B) \propto \sqrt{|B|}$
in a 2DEG and $\delta \chi_{ij} (\bm B) \propto |B|$ in a 3DEG,
which is much easier to detect experimentally
than the much weaker non-analyticities
far away from the FQPT, see Eqs.~(\ref{chi2dweak}) and (\ref{chi3dweak}). To measure local spin susceptibility one can use, for example,  ultrasensitive nitrogen-vacancy center based detectors \cite{stano_RKKY}.

\section{Conclusions}

In this paper we revisited 
FQPT in strongly interacting clean
2DEG and 3DEG with arbitrary SO splitting.
First of all, we calculated
the non-analytic corrections to the 
thermodynamic potential $\Omega$
with respect to arbitrary spin splitting
in the limit of weak electron-electron interaction.
So far, this has been done in the literature only for very limited
number of special cases 
\cite{maslov,zak1,zak2,kirk}.
Here we generalized the calculation for arbitrary 
spin splitting and arbitrary spatial dimension $D > 1$, see Eq.~(\ref{omD}).
The most important outcome of this calculation
is that even arbitrarily complex 
SO splitting is not able to cut the non-analyticity in $\Omega$ with respect to
the magnetic field, see Eq.~(\ref{sogen}).
This is a direct consequence of the 
backscattering processes shown in Fig.~\ref{fig:SOres}.
Such processes
were not taken into account
in Ref.~\cite{kirk}
where a complicated enough SO splitting 
is predicted to cut the non-analyticity.

The weak coupling regime cannot be extended 
to the phase transition point
where the electron-electron interaction is very strong.
In order to address this regime,
we treat the processes in Fig.~\ref{fig:SOres}
non-perturbatively
and find that there is a sector in the phase space
where the electron Green function exhibits
strong non-Fermi-liquid behavior.
We find that this non-Fermi-liquid sector
dominates the non-analytic response at low temperatures.
We also find that the non-analyticity 
remains negative
but it is parametrically enhanced 
from $\delta \Omega \propto - |B|^{D + 1}$
in the weak interaction limit
to $\delta \Omega \propto -|B|^{\frac{D + 3}{2}}$ in the strong coupling regime.
In particular, the non-analyticity in interacting 3DEG
is enhanced from the marginal 
$\propto B^4 \ln|B|$ term
in the weak interaction regime
to $\propto - |B|^3$ close to the FQPT
where the interaction is strong.
In addition, SO splitting makes the 
non-analytic corrections strongly anisotropic.
In particular,
if the SO splitting breaks the spin symmetry down to $\mathbb{Z}_2$, 
then the ferromagnetic ground state is necessarily Ising-like.
In this case, all soft fluctuations of the magnetic order parameter are gapped
which validates the mean-field treatment of the magnetization.
Nevertheless,
there is still a gapless fluctuation channel
originating from the resonant backscattering processes shown in Fig.~\ref{fig:SOres}(b)
which contributes to 
the negative non-analytic terms in $\Omega$ destabilizing the FQCP.
Based on our analysis, 
we conclude that the
FQPT in clean metals is always first
order regardless of the SO splitting.
The non-analytic terms leading to the instability of the FQCP in clean metals
can be measured via the non-analytic dependence
of the spin susceptibility 
on the magnetic field 
both far away, see Eqs.~(\ref{chi2dweak}) and (\ref{chi3dweak}),
or close to the FQPT, see Eqs.~(\ref{chistrong2d}) and (\ref{chistrong3d}).
The candidate materials are the pressure-tuned 3D ferromagnets ZrZn$_2$ \cite{uhlarz},
UGe$_2$ \cite{taufour}, and many others \cite{brando},
or  density-tuned 2D quantum wells \cite{hossain}.

\section*{Acknowledgments}

This work was supported by the Georg H. Endress 
Foundation, the Swiss National Science Foundation (SNSF), and NCCR SPIN.
This project received funding from the European Union's Horizon 2020 research and innovation program (ERC Starting Grant, grant agreement No 757725).

\appendix

\section{ASYMPTOTICS OF THE GREEN FUNCTION \label{app}}

We start from the Fourier representation of the Green function:
\begin{eqnarray}
&& G (\tau, \bm r) = \int \frac{d \bm p}{(2 \pi)^D} e^{i \bm p \cdot \bm r} G (\tau, \bm p) , \label{Gfou}
\end{eqnarray}
where $\tau$ is the imaginary time, 
$\bm r$  a $D$-dimensional position vector,
and $\bm p$ a $D$-dimensional momentum vector.
Here we do not indicate 
any band index because it is fixed.
The asymptotics at large $\tau \gg 1/ E_F$ and large $r = |\bm r| \gg \lambda_F$
is dominated by the vicinity of the Fermi surface $\mathcal{FS}$,
so we can expand the momentum $\bm p$ into 
the momentum $\bm k$ on the Fermi surface $\mathcal{FS}$ and 
the momentum along the outward normal $\bm n(\bm k)$ at $\bm k$,
see Fig.~\ref{fig:fsexp}(a):
\begin{eqnarray}
&& \bm p = \bm k + \bm n (\bm k) \delta p , \,\,\,\, \bm k \in \mathcal{FS} , \label{momentum}
\end{eqnarray}
where $\delta p$ is the distance from $\bm p$
to the Fermi surface $\mathcal{FS}$.
Notice that $\delta p > 0$ ($\delta p < 0$) corresponds to
the states above (below) the Fermi surface.
At large $r \gg 1/k_F$, $k_F$ is the typical momentum scale on the Fermi surface,
we have $\delta p \sim 1/r \ll k_F$,
so we can approximate the integral over $\bm p$
by the integration over a thin layer around the Fermi surface:
\begin{eqnarray}
&& G (\tau, \bm r) \approx \int\limits_{-\infty}^\infty \frac{d\delta p}{2 \pi} \int\limits_{\bm k \in \mathcal{FS}} \frac{d \bm k}{(2 \pi)^{D - 1}} \nonumber \\
&& \hspace{20pt} \times e^{i \bm k \cdot \bm r} e^{i \delta p \, \bm n(\bm k) \cdot \bm r} G (\tau, \delta p, \bm k) , \label{Gmid}\\
&& G (\tau, \delta p, \bm k) \equiv G (\tau, \bm k + \bm n(\bm k) \delta p) .
\end{eqnarray}
Here we extended the integral over $\delta p$ to the interval $(-\infty, \infty)$ because the 
convergence radius of this integral is very short at $r \to \infty$, namely $\delta p \sim 1/r$.
Hence, we approximated the initial Fourier 
transform Eq.~(\ref{Gfou})
by the integral over the  fiber bundle
$\mathcal{FS} \times (-\infty, \infty)$.

As $r \to \infty$, we can use the stationary phase method to find the asymptotics.
First, we evaluate the integral over the Fermi surface.
The stationary condition for the phase of the rapidly oscillating factor $e^{i \bm k \cdot \bm r}$ in Eq.~(\ref{Gmid})
reads:
\begin{eqnarray}
&& d \bm k \cdot \bm r = 0 ,
\end{eqnarray}
where $d \bm k$ is an arbitrary infinitesimal (but non-zero) element 
of the tangent space $\mathcal{T}(\bm k)$ attached to the Fermi surface $\mathcal{FS}$ at the point $\bm k$.
Hence, $\bm r$ has to be orthogonal to the 
whole linear space $\mathcal{T}(\bm k)$
which has  codimension one.
This means that the stationary phase condition is satisfied 
at the points $\bm k' \in \mathcal{FS}$
where $\bm r$ is collinear with the normals $\bm n(\bm k')$:
\begin{eqnarray}
&& \bm n(\bm k') = s(\bm k', \bm n_{\bm r}) \bm n_{\bm r} , \,\,\,\, \bm n_{\bm r} = \frac{\bm r}{r} , \label{norm}
\end{eqnarray}
where $s(\bm k', \bm n_{\bm r}) = +1$ ($s(\bm k', \bm n_{\bm r}) = -1$) 
if the outward normal $\bm n(\bm k')$ and
the radius vector $\bm r$ 
are parallel (anti-parallel).
We include all such points $\bm k'$
into a set $\mathcal{P} (\bm n_{\bm r})$:
\begin{eqnarray}
&& \mathcal{P} (\bm n_{\bm r}) = \{\bm k' \in \mathcal{FS} |\bm n(\bm k') = \pm \bm n_{\bm r}\} .
\end{eqnarray}
It is clear that $\mathcal{P}(-\bm n_{\bm r}) = \mathcal{P}(\bm n_{\bm r})$
and $s(\bm k', -\bm n_{\bm r}) = -s(\bm k', \bm n_{\bm r})$.

At this point we can take the integral over $\delta p$ in Eq.~(\ref{Gmid}):
\begin{eqnarray}
&& \hspace{-15pt} \int\limits_{-\infty}^\infty \frac{d\delta p}{2 \pi} e^{i \delta p \, \bm n(\bm k') \cdot \bm r} G (\tau, \delta p, \bm k')  = \mathcal{G} (\tau, s(\bm k', \bm n_{\bm r}) r , \bm k') , \label{G1d}
\end{eqnarray}
where we used that $\bm k' \in \mathcal{P} (\bm n_{\bm r})$
and $s(\bm k', \bm n_{\bm r}) = \bm n(\bm k') \cdot \bm n_{\bm r} = \pm 1$, see Eq.~(\ref{norm}).
Here $\mathcal{G} (\tau, x, \bm k)$ is the 1D Fourier transform of the Green function:
\begin{eqnarray}
&& \mathcal{G} (\tau, x, \bm k) = \int\limits_{-\infty}^\infty \frac{d\delta p}{2 \pi} e^{i \delta p \, x} G (\tau, \delta p, \bm k) , \,\, \bm k \in \mathcal{FS} .
\end{eqnarray}
Note that such a 1D Fourier transform is generally dependent on 
the point $\bm k \in \mathcal{FS}$.
Substituting this into Eq.~(\ref{Gmid}), we find:
\begin{eqnarray}
&& \hspace{-15pt} G (\tau, \bm r) \approx \sum\limits_{\bm k' \in \mathcal{P}(\bm n_{\bm r})} e^{i \bm k' \cdot \bm r} J_{\bm k'} (r)   \mathcal{G} (\tau, s(\bm k', \bm n_{\bm r}) r, \bm k') , \label{Gmid2} \\
&& \hspace{-15pt} J_{\bm k'} (r) = \int\limits_{\bm k \in \mathcal{FS}} \frac{d \bm k}{(2 \pi)^{D - 1}} e^{i (\bm k - \bm k') \cdot \bm r} . \label{J}
\end{eqnarray}
The function $J_{\bm k'} (r)$ appears due to the integration over a small vicinity of a point $\bm k' \in \mathcal{P} (\bm n_{\bm r})$.

The integral $J_{\bm k'}(r)$ converges due to the finite curvature of the Fermi surface at $\bm k'$.
This is true even if the interaction is very strong such that the Fermi surface becomes critical.
In order to evaluate this integral,
it is convenient to introduce an auxiliary function
$\varepsilon (\bm p)$ with the following properties:
\begin{eqnarray}
&& \varepsilon (\bm p) = 0 \,\, {\rm if} \,\, \bm p \in \mathcal{FS} , \label{FS}\\
&& \bm v(\bm p) = \frac{\partial \varepsilon (\bm p)}{\partial \bm p} \ne 0 \,\, {\rm if} \,\, \bm p \in \mathcal{FS} . \label{v}
\end{eqnarray}
We notice here that there are infinitely many choices for such a function but
we will show that the result is independent 
of the choice.
In case of a free electron gas the natural choice for $\varepsilon (\bm p)$ is the electron dispersion.
The condition Eq.~(\ref{v}) is required to generate the outward normal $\bm n(\bm k)$ at each point $\bm k \in \mathcal{FS}$:
\begin{eqnarray}
&& \bm v(\bm k) = v(\bm k) \bm n(\bm k) , \,\, \bm k \in \mathcal{FS} . \label{vel}
\end{eqnarray}
Here $v(\bm k) \ne 0$ for all $\bm k \in \mathcal{FS}$.

Consider two close points $\bm k \in \mathcal{FS}$ 
and $\bm k' \in \mathcal{FS}$.
According to Eq.~(\ref{FS}), 
we can write:
\begin{eqnarray}
&& \varepsilon (\bm k) = \varepsilon (\bm k') = 0 .
\end{eqnarray}
Using the Taylor series expansion,
we find:
\begin{eqnarray}
&& 0 = \varepsilon (\bm k) \approx \varepsilon (\bm k') + (\bm k - \bm k') \cdot \bm v(\bm k') \nonumber \\
&& + \frac{1}{2} (\bm k - \bm k')^T R(\bm k') (\bm k - \bm k') , \label{eps}\\
&& R_{ij}(\bm p) = \frac{\partial^2 \varepsilon(\bm p)}{\partial p_i \partial p_j} , \label{R}
\end{eqnarray}
where we used the matrix notations in Eq.~(\ref{eps}), and the superscript
 $^T$ stands for transposition.
Substituting Eq.~(\ref{vel}) into Eq.~(\ref{eps}),
we find:
\begin{eqnarray}
&& (\bm k - \bm k') \cdot \bm n(\bm k') \approx -  (\bm k - \bm k')^T \frac{R(\bm k')}{2 v(\bm k')} (\bm k - \bm k') . \label{curv}
\end{eqnarray}
In Eq.~(\ref{J}) $\bm k' \in \mathcal{P}(\bm n_{\bm r})$,
so $\bm n(\bm k')$ satisfies Eq.~(\ref{norm}).
This allows us to write:
\begin{eqnarray}
&& \hspace{-20pt} (\bm k - \bm k') \cdot \bm r \approx - s(\bm k', \bm n_{\bm r}) r  (\bm k - \bm k')^T \frac{R(\bm k')}{2 v(\bm k')} (\bm k - \bm k') . \label{quad}
\end{eqnarray}
The expression is quadratic with respect to the 
small difference $\bm k - \bm k'$.
At large $r$ the convergence radius of the integral $J_{\bm k'}(r)$ 
scales as $|\bm k - \bm k'| \propto 1/\sqrt{r}$,
so we indeed have to integrate over a small 
vicinity of $\bm k' \in \mathcal{P}(\bm n_{\bm r})$
and the Taylor expansion is valid.
Equation~(\ref{quad}) also 
shows that the component of $\bm k - \bm k'$ 
along the normal $\bm n(\bm k')$ is only quadratic with respect to 
$\bm k - \bm k'$,
so at a given accuracy we can approximate
$\bm k - \bm k'$ on the right-hand side of Eq.~(\ref{quad})
by its orthogonal projection onto the 
tangent space $\mathcal{T}(\bm k')$:
\begin{eqnarray}
&&\hspace{-7pt} J_{\bm k'}(r) \approx \int\limits_{\kap \in \mathcal{T}(\bm k')} \frac{d \kap}{(2 \pi)^{D - 1}} e^{-i r s(\bm k', \bm n_{\bm r}) \kap^T A(\bm k') \kap} , \label{gauss} \\
&&\hspace{-7pt} A(\bm k') = \frac{R_{\mathcal{T}}(\bm k')}{2 v(\bm k')}  ,
\end{eqnarray}
where $R_{\mathcal{T}}(\bm k')$ is the 
restriction of the tensor $R (\bm k')$ to the tangent space $\mathcal{T}(\bm k')$.
We reduced the initial integral $J_{\bm k'}(r)$ to a standard Gaussian integral:
\begin{eqnarray}
&& J_{\bm k'} (r) \approx \left(\frac{1}{4 \pi r}\right)^{\frac{D - 1}{2}} \frac{e^{-i \frac{\pi}{4} s(\bm k', \bm n_{\bm r}) \mathfrak{S}(A(\bm k'))}}{\sqrt{|\det A(\bm k')|}} , \\
&& \mathfrak{S}(A(\bm k')) = \sum\limits_{i = 1}^{D - 1} {\rm sgn}(a_i) ,
\end{eqnarray}
where $a_i$ are all $D - 1$ eigenvalues of the
symmetric matrix $A(\bm k')$.
In all cases that we consider in this paper
$\det A(\bm k) \ne 0$ at any point $\bm k \in \mathcal{FS}$.
In other words, we consider only Fermi surfaces 
with non-zero Gauss curvature at each point.

Let us check
that the matrix $A(\bm k)$, $\bm k \in \mathcal{FS}$,
is indeed independent of the choice of $\varepsilon (\bm p)$.
For this, we consider another parametrization:
\begin{eqnarray}
&& \tilde \varepsilon (\bm p) = f(\bm p) \varepsilon(\bm p) ,
\end{eqnarray}
where $f(\bm p)$ is an arbitrary smooth function such that $f(\bm p) \ne 0$.
As $f(\bm p) \ne 0$, then $\tilde \varepsilon (\bm p) = 0$ if and only if 
$\varepsilon(\bm p) = 0$, i.e. 
$\tilde \varepsilon (\bm p) = 0$ defines the Fermi surface $\mathcal{FS}$.
Let us find the velocity $\tilde v(\bm k)$ when $\bm k \in \mathcal{FS}$:
\begin{eqnarray}
&& \hspace{-20pt} \tilde v(\bm k) \bm n(\bm k) =  \frac{\partial \tilde \varepsilon (\bm k)}{\partial \bm k} = \frac{\partial f(\bm k)}{\partial \bm k} \varepsilon(\bm k)  + f(\bm k) v(\bm k) \bm n (\bm k) ,
\end{eqnarray}
where $v(\bm k)$ is defined in Eqs.~(\ref{v}) and (\ref{vel}).
As $\bm k \in \mathcal{FS}$, then $\varepsilon (\bm k) = 0$, so we find:
\begin{eqnarray}
&& \tilde v(\bm k) = f(\bm k) v(\bm k), \,\, \bm k \in \mathcal{FS} . \label{vtil}
\end{eqnarray}
Similarly, we can calculate the tensor $\tilde R(\bm k)$, $\bm k \in \mathcal{FS}$:
\begin{eqnarray}
&& \tilde R_{ij}(\bm k) = \frac{\partial^2 \tilde \varepsilon (\bm k)}{\partial k_i \partial k_j} = f(\bm k) R_{ij}(\bm k) \nonumber \\
&& \hspace{0pt} + v(\bm k) \left(\frac{\partial f}{\partial k_i} \bm n_j(\bm k) + \frac{\partial f}{\partial k_j} \bm n_i(\bm k)\right) , \,\, \bm k \in \mathcal{FS} , \label{Rtil}
\end{eqnarray}
where $R_{ij}(\bm k)$ is defined in Eq.~(\ref{R}).
Second line in Eq.~(\ref{Rtil})
contains a term which vanishes in all 
products $\kap^T \tilde R(\bm k) \kap$ where $\kap \in \mathcal{T}(\bm k)$ i.e. $\kap \cdot \bm n(\bm k) = 0$.
This means that the restriction to the tangent space $\mathcal{T}(\bm k)$ is especially simple:
\begin{eqnarray}
&& \tilde R_{\mathcal{T}}(\bm k) = f(\bm k) R_{\mathcal{T}} (\bm k) , \,\, \bm k \in \mathcal{FS} . \label{rtil}
\end{eqnarray}
Using Eqs.~(\ref{vtil}) and (\ref{rtil}),
we indeed find that the operator $A(\bm k)$, $\bm k \in \mathcal{FS}$,
is invariant with respect to different choices of $\varepsilon(\bm p)$:
\begin{eqnarray}
&& \tilde A(\bm k) = \frac{\tilde R_{\mathcal{T}}(\bm k)}{2 \tilde v(\bm k)} = \frac{R_{\mathcal{T}}(\bm k)}{2 v(\bm k)} = A(\bm k) , \,\, \bm k \in \mathcal{FS} .
\end{eqnarray}

All in all, the long range asymptotics of the Green function of an interacting Fermi gas
is the following:
\begin{eqnarray}
&& G (\tau, \bm r) \approx \sum\limits_{\bm k' \in \mathcal{P} (\bm n_{\bm r})} \frac{C(\bm k', \bm n_{\bm r})}{(4 \pi r)^{\frac{D-1}{2}}} e^{i \bm k' \cdot \bm r} \nonumber \\
&& \hspace{20pt} \times \mathcal{G} (\tau, s(\bm k', \bm n_{\bm r}) r, \bm k') , \label{Gasympt} \\
&& C(\bm k', \bm n_{\bm r}) = \frac{e^{-i \frac{\pi}{4} s(\bm k', \bm n_{\bm r}) \mathfrak{S}(A(\bm k'))}}{\sqrt{|\det A(\bm k')|}} . \label{Ck}
\end{eqnarray}
This is the general result which is suitable for a Fermi surface of arbitrary geometry.
Next, we give examples for spherical or nearly spherical Fermi surfaces.

\subsection{Spherical Fermi surface}

The simplest example is the spherical Fermi surface with the Fermi momentum $k_F$.
We considered this case in Ref.~\cite{miserev}.
For an arbitrary direction $\bm n_{\bm r}$
there are exactly two points on the Fermi surface
whose normals are collinear with $\bm n_{\bm r}$:
\begin{eqnarray}
&& \mathcal{P}(\bm n_{\bm r}) = \{\pm k_F \bm n_{\bm r}\} .
\end{eqnarray}
In this case, the sum over $\bm k'$ in Eq.~(\ref{Gasympt})
contains only two terms, namely,
$\bm k' = \pm k_F \bm n_{\bm r}$.
In order to calculate the matrix $A(\bm k')$,
we consider the function:
\begin{eqnarray}
&& \varepsilon (\bm p) = \frac{p^2 - k_F^2}{2} .
\end{eqnarray}
The velocity $v(\bm p)$ and the tensor $R(\bm p)$ are then the following:
\begin{eqnarray}
&& v(\bm p) = p , \\
&& R_{ij}(\bm p) = \delta_{ij} .
\end{eqnarray}
This allows us to identify the matrix $A(\bm k)$, $|\bm k| = k_F$:
\begin{eqnarray}
&& A(\bm k) = \frac{I}{2 k_F} , \,\, |\bm k| = k_F ,
\end{eqnarray}
where $I$ is the $(D - 1) \times (D - 1)$ 
identity matrix on the tangent space $\mathcal{T}(\bm k)$.
Substituting this into Eq.~(\ref{Gasympt}),
we find the asymptotics of the Green function
in case of the spherical Fermi surface:
\begin{eqnarray}
&& G(\tau, \bm r) \approx \left(\frac{k_F}{2\pi r}\right)^{\frac{D - 1}{2}} \nonumber \\
&& \hspace{5pt} \times \left(e^{i (k_F r - \vartheta)} \mathcal{G}(\tau, r) + e^{-i (k_F r - \vartheta)} \mathcal{G}(\tau, -r)\right) , \label{Gspher} \\
&& \vartheta = \frac{\pi}{4}(D - 1) , \label{vartheta} \\
&& \mathcal{G}(\tau, x) = \int\limits_{-\infty}^\infty \frac{d\delta p}{2 \pi} e^{i \delta p \, x} G(\tau, \delta p) . \label{Gs1d}
\end{eqnarray}
Here we also used the spherical symmetry, i.e. 
$G(\tau, \bm p) = G(\tau, p)$,
so $G(\tau, \delta p, \bm k)$ is independent of $\bm k \in \mathcal{FS}$.

\subsection{Nearly spherical Fermi surface}

Here we consider another example when the Fermi surface is nearly spherical
and can be modeled by the following dispersion:
\begin{eqnarray}
&& \varepsilon(\bm p) = \frac{p^2 - k_F^2}{2 m} - \beta(\bm p) .
\end{eqnarray}
We denote points on the Fermi surface $\mathcal{FS}$
by $\bm k$, they satisfy the equation $\varepsilon(\bm k) = 0$:
\begin{eqnarray}
&& k^2 = k_F^2 + 2 m \beta (\bm k) . \label{fermimom}
\end{eqnarray}
In this part we make the following assumptions about the smooth function
$\beta(\bm p)$:
\begin{eqnarray}
&& |\beta(\bm k)| \ll E_F , \,\,\, \nabb \beta(\bm k) \equiv \frac{\partial \beta(\bm p)}{\partial \bm p}\left.\vphantom{\frac{\partial \beta(\bm p)}{\partial \bm p}} \right|_{\bm p = \bm k} \ll v_F , \label{cond}
\end{eqnarray}
where $\bm k \in \mathcal{FS}$,
$2 E_F = k_F v_F$ is the Fermi energy, and
$v_F = k_F/m$  the Fermi velocity.
Using the first condition in Eq.~(\ref{cond}),
we find the approximate Fermi surface equation:
\begin{eqnarray}
&& k (\bm e)  \approx k_F + \frac{\beta(\bm e)}{v_F} , \,\, k(\bm e) \bm e \in \mathcal{FS} ,  \label{fseq}
\end{eqnarray}
where $\bm e$ is an arbitrary unit vector
and $\beta(\bm e)$ stands for $\beta(k_F \bm e)$.

The outward normal $\bm n(\bm k)$ at $\bm k \in \mathcal{FS}$
is defined though the gradient 
of $\varepsilon (\bm p)$ at $\bm p = \bm k$:
\begin{eqnarray}
&& v(\bm k) \bm n(\bm k) = \frac{\bm k}{m} - \nabb \beta(\bm k)  , \label{nk}\\
&& v^2(\bm k) = \frac{k^2}{m^2} - 2 \frac{\bm k \cdot \nabb \beta(\bm k)}{m} + \left(\nabb \beta(\bm k)\right)^2 , \label{vk}
\end{eqnarray}
where the second equation here is just the first one squared.
Here is where we use the second condition in Eq.~(\ref{cond}).
In linear order in $\beta(\bm k)$ we find:
\begin{eqnarray}
&& v(\bm k) \approx \frac{k(\bm n)}{m} - \bm n \cdot \nabb \beta(\bm n) , \label{vn}\\
&& \bm k \cdot \bm n(\bm k) \approx k(\bm n) = k_F + \frac{\beta(\bm n)}{v_F} , \label{kn}
\end{eqnarray}
where $\beta(\bm n)$ stands for $\beta (k_F \bm n)$.

We are only interested in the points $\bm k' \in \mathcal{FS}$ 
with normals $\bm n (\bm k') = s \bm n_{\bm r}$, $s = \pm 1$.
Using Eq.~(\ref{kn}), we find the oscillating phase:
\begin{eqnarray}
&& \bm k' \cdot \bm r = s r \bm k' \cdot \bm n(\bm k') \approx s k(s \bm n_{\bm r}) r , \label{phase}
\end{eqnarray}
where we used Eq.~(\ref{kn}).
Neglecting the weak dependence of the prefactor $C(\bm k', \bm n_{\bm r})$ on $\beta(\bm k')$,
see Eq.~(\ref{Ck}),
we find the asymptotic behavior of the Green function in case of a nearly spherical Fermi surface:
\begin{eqnarray}
&& \hspace{-20pt} G(\tau, \bm r) \approx \left(\frac{k_F}{2\pi r}\right)^{\frac{D - 1}{2}} \nonumber \\
&& \hspace{-20pt} \times \left(e^{i (k(\bm n_{\bm r}) r - \vartheta)} \mathcal{G}(\tau, r) + e^{-i (k(-\bm n_{\bm r}) r - \vartheta)} \mathcal{G}(\tau, -r)\right) , \label{Gnearspher} 
\end{eqnarray}
where $k(\bm n)$ is given by Eq.~(\ref{fseq})
and $\vartheta$ by Eq.~(\ref{vartheta}).
Here, $\mathcal{G}(\tau, x)$ is calculated at $\beta(\bm p) = 0$, i.e. it coincides with 
the spherically symmetric case.
Importantly, the oscillatory factors
$e^{\pm i k(\pm \bm n_{\bm r}) r}$ 
in Eq.~(\ref{Gnearspher}) depend explicitly on
$\beta(\pm \bm n_{\bm r})$,
which is crucial for the resonant scattering processes near the Fermi surface.

\end{document}